\documentclass[useAMS,usenatbib]{mnras}

\usepackage{epsfig}
\usepackage{aas_macros}
\usepackage{natbib}
\usepackage{times}
\usepackage{graphicx, bm, amssymb, xcolor}
\usepackage{verbatim}
\usepackage[all]{hypcap}
\usepackage{xspace}
\usepackage{multirow}
\usepackage{amsmath}
\usepackage{epstopdf}

\newcommand{\msun}{\mbox{M$_\odot$}}
\newcommand{\yr}{\mbox{${\rm yr}$}}
\newcommand{\myr}{\mbox{${\rm Myr}$}}
\newcommand{\gyr}{\mbox{${\rm Gyr}$}}
\newcommand{\pc}{\mbox{${\rm pc}$}}
\newcommand{\kpc}{\mbox{${\rm kpc}$}}
\newcommand{\kms}{\mbox{${\rm km}~{\rm s}^{-1}$}}

\newcommand{\be}{\begin{equation}}
\newcommand{\ee}{\end{equation}}
\newcommand{\bea}{\begin{eqnarray}}
\newcommand{\eea}{\end{eqnarray}}

\title{A unified model for the maximum mass-scales of molecular clouds, stellar clusters, and high-redshift clumps}
\author{Marta~Reina-Campos\thanks{reina.campos@uni-heidelberg.de} and J.~M.~Diederik~Kruijssen\thanks{kruijssen@uni-heidelberg.de}\\
Astronomisches Rechen-Institut, Zentrum f\"{u}r Astronomie der Universit\"{a}t Heidelberg, M\"{o}nchhofstra\ss e 12-14, 69120 Heidelberg, Germany}

\begin{document}

\date{Accepted 2017 March 27. Received 2017 March 27; in original form 2016 December 21.}

\pagerange{\pageref{firstpage}--\pageref{lastpage}} \pubyear{2016}

\maketitle

\label{firstpage}

\begin{abstract}
We present a simple, self-consistent model to predict the maximum masses of giant molecular clouds (GMCs), stellar clusters and high-redshift clumps as a function of the galactic environment. Recent works have proposed that these maximum masses are set by shearing motions and centrifugal forces, but we show that this idea is inconsistent with the low masses observed across an important range of local-Universe environments, such as low-surface density galaxies and galaxy outskirts. Instead, we propose that feedback from young stars can disrupt clouds before the global collapse of the shear-limited area is completed. We develop a shear-feedback hybrid model that depends on three observable quantities: the gas surface density, the epicylic frequency, and the Toomre parameter. The model is tested in four galactic environments: the Milky Way, the Local Group galaxy M31, the spiral galaxy M83, and the high-redshift galaxy zC406690. We demonstrate that our model simultaneously reproduces the observed maximum masses of GMCs, clumps and clusters in each of these environments. We find that clouds and clusters in M31 and in the Milky Way are feedback-limited beyond radii of 8.4 and 4 kpc, respectively, whereas the masses in M83 and zC406690 are shear-limited at all radii. In zC406690, the maximum cluster masses decrease further due to their inspiral by dynamical friction. These results illustrate that the maximum masses change from being shear-limited to being feedback-limited as galaxies become less gas-rich and evolve towards low shear. This explains why high-redshift clumps are more massive than GMCs in the Local Universe.
\end{abstract}


\section{Introduction} \label{sec:intro}

A large body of recent work has demonstrated that star formation in galaxies is correlated with the molecular gas \citep[e.g.][]{bigiel08,leroy08,schruba11,leroy13,tacconi13}. \citet{solomon87} observed that the bulk of molecular gas in the Milky Way resides in the most massive giant molecular clouds (GMCs). Following from the correlation between the star formation rate (SFR) and the molecular gas, the most massive GMCs also contain most of the star formation activity in the galaxy \citep[e.g.][]{murray11}. Such star formation `hubs' are observed to produce dense stellar clusters in the Universe \citep[e.g.][]{longmore14}. It is unknown how these massive structures form, or what sets their extreme mass-scale.

The most massive, high-redshift ($z\sim1$--$3$) star-forming galaxies exhibit a more clumpy structure than their Local-Universe counterparts. The giant clumps have typical observed masses of $\sim 10^8$--$10^9\,\msun$, but they can account for $\sim10$--$20$ per cent of the SFR of their host galaxy (e.g.~\citealt{genzel11}, \citealt{oklopcic17}, \citealt{soto17}).
\citet{oklopcic17} use the FIRE simulations to study the giant clumps that dominate the morphology of massive high-redshift galaxies. They find that the high-redshift molecular clumps have similar star formation efficiencies and follow the same mass-size relations as local-Universe molecular clouds, so they suggest that the giant clumps might be analogous to GMCs in the Local Universe.

While the formation physics of the most massive clusters are still poorly understood, be it young massive clusters (YMCs) or globular clusters, they represent an ideal opportunity to probe the most extreme forms of star formation. A popular hypothesis that is currently being tested is that globular clusters may be the high-redshift equivalent to the YMCs forming in the Local Universe that have survived for a Hubble time (e.g.~\citealt{kravtsov05}, \citealt{elmegreen10}, \citealt{shapiro10}, \citealt{kruijssen15b}). If this hypothesis holds, then an understanding of the physics setting the maximum cluster mass-scales would enable the use of globular clusters as tracers of the star-forming conditions in high-redshift environments.

The characteristic mass-scale of the GMC mass function has been observed to depend on the hierarchical structure of the interstellar medium (ISM) and it is known to vary with the galactic environment (\citealt{hughes13}, \citealt{colombo14}, \citealt{freeman17}). Likewise, the young cluster mass function follows an exponentially truncated power-law with index -2 (i.e.~a \citealt{schechter76} function, e.g.~\citealt{gieles06a}, \citealt{larsen09}, \citealt{bastian12}, \citealt{konstantopoulos13}), with values for the characteristic truncation mass $M_{\rm c}$ in the range $0.1 - 50 \times 10^5\,\msun$. Moreover, the recovered truncation mass has been found to change as a function of galactocentric radius within the same galaxy by \citet{adamo15b}. As for GMCs, the observed variation suggests a dependence on the galactic environment. If stellar clusters form in the overdensities of the GMCs, one could expect a certain correspondence between the mass function of both objects, but cloud fragmentation and hierarchical cluster growth make this mapping non-trivial \citep[e.g.][]{longmore14}. A possible way to overcome this problem is to only consider the maximum GMC and cluster mass-scales, because these must correspond to the optimal case of minimal fragmentation and maximal hierarchical growth.

\citet{kruijssen14c} suggested that the maximum GMC mass and the maximum stellar cluster mass might have a common origin and it may correspond to the maximum mass that could collapse against centrifugal forces, i.e.~the Toomre mass. \citet{toomre64} considers whether shear due to the rotation of the disc and the internal kinetic energy of the gas are sufficient to stop the collapse due to a gravitational instability. \citet{toomre64} proposes that such shearing motions and centrifugal forces are the mechanisms setting the maximum region for collapse and derives the maximum size of the collapsing region, called Toomre length. For a given gas surface density, this length-scale directly provides the Toomre mass. This definition of a shear-limited mass-scale has recent been put forward to explain the maximum mass of masses of GMCs, stellar clusters, and high-redshift clumps, implying that shear is the only limiting factor mechanism \citep[e.g.][]{dekel09,kruijssen14c,adamo15b,freeman17}. However, this idea assumes that the collapse of the shear-limited area can proceed fast enough to condense into a single object. In this paper, we show that, across an important part of parameter space, feedback from young stars disrupts the cloud before the global collapse of the shear-limited area is completed. 

The idea that stellar feedback can end up destroying the molecular cloud in which stars form has been put forward several decades ago \citep[e.g.][]{oort55,larson81}. In this work, we add the idea of cloud destruction by feedback to the classical Toomre approach and derive a self-consistent, simple model to simultaneously predict the maximum masses of molecular clouds and stellar clusters, from local galaxies out to the clumpy, star-forming systems observed at high redshift. We show that the feedback time (i.e.~the time it takes for the stellar feedback to destroy the cloud) can be smaller than the free-fall time of the shear-limited region. In such a situation, the collapsed mass is smaller than the shear-limited Toomre mass. The goal of this model is to shed light on the conditions under which the densest structures in galaxies form, and how the local-Universe GMCs and stellar clusters may be connected to their high-redshift analogues of giant molecular clumps and globular clusters.

The structure of this paper is as follows. We present a self-consistent simple model to simultaneously predict the maximum masses of molecular clouds and stellar clusters, from local galaxies out to the clumpy, star-forming systems observed at high redshift. We introduce the idea that the feedback time (i.e. time it takes to the stellar feedback to destroy the cloud) can be smaller than the two-dimensional free-fall time of the shear-limited region. In such a scenario, the collapsed mass will be smaller than the shear-limited Toomre mass. We present the derivation of our model in Section~\ref{sec:model}. Then we predict the maximum cloud and cluster mass for a region of the parameter space (Section~\ref{sec:results}). We then test the predictions of our model for the Milky Way, the Local Group disc galaxy M31, the grand-design spiral galaxy M83, and the high-redshift galaxy zC406690 to their observed maximum mass-scales and present the results in Section~\ref{sec:testmethod}. Finally, we present the conclusions of this work in Section~\ref{sec:concl}.

\section{Model} \label{sec:model}

In this section we derive a simple analytical model that self-consistently predicts the maximum mass-scales of GMCs and stellar clusters, from local galaxies out to high redshift. We consider the situation in which the mass assembly of clouds and clusters can be limited by shear and feedback. Which of these two mechanisms ends up setting the maximum mass-scales depends on whether or not the collapse of the GMC enclosed by the shear-limited area proceeds more rapidly (i.e.~the free-fall time) than the time it takes stellar feedback to disperse the GMC (i.e.~the feedback time).

\subsection{Mass of the most massive GMC}

We first determine the collapse time of a GMC enclosed by the shear-limited area. In a differentially rotating disc, \citet{toomre64} explores the possibility of compensating a gravitational instability with internal kinetic energy and shear. The author finds that the random motions set a minimum scale for collapse that corresponds to the Jeans length $\lambda_{\rm J}$, whereas the maximum scale is set by shear and corresponds to the Toomre length $\lambda_{\rm T}$, which is set by galactic-scale quantities,

\be
\lambda_{\rm T} = \dfrac{4\pi^2 G\Sigma_{\rm g}}{\kappa^{2}},
\label{eq:toomrelength}
\ee
\noindent

where $G$ is the gravitational constant, $\Sigma_{\rm g}$ is the gas surface density and $\kappa$ corresponds to the epicyclic frequency. The disc is stable to perturbations with wavelengths larger than this length-scale. Toomre posits that collapse can only take place when $\lambda_{\rm J}<\lambda_{\rm T}$. Given that the Jeans length depends on local conditions and increases with the local velocity dispersion, the turbulent energy will dissipate even if initially $\lambda_{\rm J}>\lambda_{\rm T}$, thus decreasing the Jeans length and eventually meeting the unstability condition. The collapse then occurs on a scale $\lambda_{\rm T}$, which is therefore the largest scale on which collapse can take place in our model.

The collapse length-scale can naturally be related to an equivalent mass-scale (i.e.~the Toomre mass) for a given gas surface density:
\be
M_{\rm T} = \pi\Sigma_{\rm g} \dfrac{\lambda_{\rm T}^2}{4} = \dfrac{4\pi^5G^2\Sigma_{\rm g}^3}{\kappa^4}.
\label{eq:toomremass}
\ee
\noindent

The characteristic time-scale associated to the collapse of a region enclosed by shear can be determined as the two-dimensional free-fall time of the sheet of gas derived by \citet{burkert04}. They assume finite, self-gravitating sheets of gas and derive the typical infall time for a subregion within a collapsing sheet of radius $r$. Using numerical simulations, they demonstrate that this time-scale also describes the time it takes to the edges of the collapsing sheet to reach the center. Using that our maximum collapsing region has a size $r = \lambda_{\rm T}/2$ and equation~(\ref{eq:toomrelength}) to describe the Toomre length, we obtain a collapsing time-scale $t_{\rm ff,2D}$ inversely proportional to the epicyclic frequency,

\be
t_{\rm ff, 2D} = \sqrt{\dfrac{r}{\pi G\Sigma_{\rm g}}} = \sqrt{\dfrac{\lambda_{\rm T}}{2\pi G\Sigma_{\rm g}}} = \dfrac{\sqrt{2\pi}}{\kappa}.
\label{eq:tff-2D}
\ee
\noindent

In environments of low gas surface density and low shear (i.e.~low $\lambda_{\rm T}$), the two-dimensional free-fall time may become larger than the time it takes massive stars to destroy the cloud (the feedback time-scale $t_{\rm fb}$). In the case where $t_{\rm fb} < t_{\rm ff, 2D}$, the collapsing GMC will be destroyed by stellar feedback before it has finished the collapse, so the maximum mass of the collapsing region will be limited by the minimum time-scale.

We use equation~(20) in \citet{kruijssen12d} to determine the feedback time-scale, that is, the time needed to achieve pressure balance between the feedback energy density and the turbulent pressure of the ISM, by setting ambient gas density equal to the mid-plane density. This leads to:
\be
t_{\rm fb} = \dfrac{t_{\rm sn}}{2} \left(1+\sqrt{1+\dfrac{4\pi^2G^2t_{\rm ff,g} Q^2\Sigma_{\rm g}^2}{\phi_{\rm fb}\epsilon_{\rm ff}t_{\rm sn}^2\kappa^2}}\right),
\label{eq:tfbgmc}
\ee
\noindent
where $t_{\rm sn} = 3\,\myr$ is the typical time of the first supernova explosion,$t_{\rm ff,g}$ is the free-fall time of the ISM, $Q$ is the Toomre parameter, $\phi_{\rm fb} \approx 0.16\,\rm cm^2\,\rm s^{-3}$ is a constant that represents the rate at which feedback injects energy into the ISM per unit stellar mass, and $\epsilon_{\rm ff}$ corresponds to the star formation efficiency per free-fall time. We use the empirically motivated assumption that $\epsilon_{\rm ff}$ is approximately constant from \citet{elmegreen02}, $\epsilon_{\rm ff} = 0.012$ (also see \citealt{krumholz12}). Eq.~(\ref{eq:tfbgmc}) shows that in high-surface density environments, it may take time for feedback to overcome the ambient pressure, whereas in low-surface density environments the feedback time-scale can be shorter than global, two-dimensional free-fall time. This definition of the feedback time assumes that stars can start forming immediately once a cloud starts to collapse, which is justified because the hierarchical structure of the interstellar medium means that locally the free-fall time is much shorter than it is globally, enabling star formation in local overdensities once the GMC has condensed out of the background.

The \citet{toomre64} $Q$ parameter is defined as:
\be
Q \equiv \dfrac{\kappa\sigma}{\pi G\Sigma_{\rm g}},
\label{eq:toomreQ}
\ee
\noindent
where $\sigma$ corresponds to the one-dimensional velocity dispersion of the gas. This parameter is used to asses the stability of the disc against shear. The case $\lambda_{\rm J} = \lambda_{\rm T}$ refers to the value of $Q = 2$, which implies marginal stability. A region of the disc will collapse if its value of the parameter is $Q<2$, which implies $\lambda_{\rm J}<\lambda_{\rm T}$\footnote{This differs from the classical $Q=1$ condition, because we are interested in the largest unstable scale rather than the most unstable scale.}. 

\begin{figure}
	\centering
	\includegraphics[width=0.6\hsize]{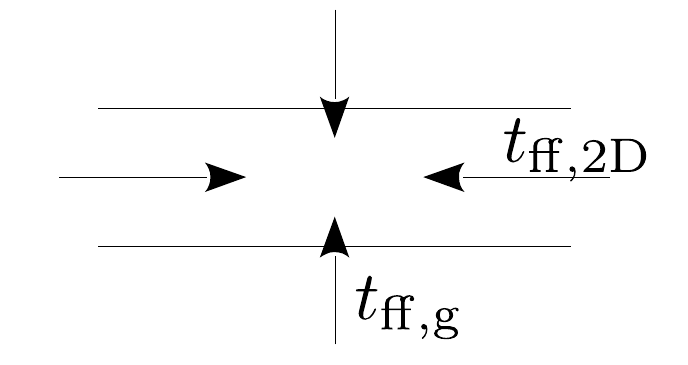}
	\caption{
		\label{fig:fig0-scheme-collapse}
		Schematic representation of the collapse. The two-dimensional collapsing time-scale $t_{\rm ff,2D}$ describes the collapse proceeding on the horizontal direction, whereas the vertical mid-plane free-fall time $t_{\rm ff, g}$ characterises the gravitational collapse towards star formation, and thus, the feedback time-scale $t_{\rm fb}$.
	}
\end{figure}

The free-fall time of the ISM $t_{\rm ff,g}$ corresponds to the collapsing time in the vertical direction at the mid-plane density as represented in Fig.~\ref{fig:fig0-scheme-collapse} and it characterises the gravitational collapse towards star formation,
\be
t_{\rm ff, g} = \sqrt{\dfrac{3\pi}{32G\rho_{\rm g}}}.
\label{eq:tff-g}
\ee
where $\rho_{\rm g}$ corresponds to the mid-plane density of the ISM and can be defined following \citet{krumholz05} and assuming a disc in hydrostatic equilibrium as:
\be
\rho_{\rm g} = \dfrac{\phi_{\rm P} \kappa^2}{2\pi Q^2 G},
\ee
\noindent
where $\phi_{\rm P} = 3$ is a constant to account for the gravity of the stars. The hydrostatic equilibrium assumption has the major advantage that the properties of the ISM can be described in terms of just three variables: the gas surface density $\Sigma_{\rm g}$, the epicyclic frequency $\kappa$ and the stability parameter $Q$.

Using equation~(\ref{eq:toomremass}) and~(\ref{eq:tff-2D}), we can determine the dependence of the Toomre mass in the collapsing time-scale as $M_{\rm T} \propto t_{\rm ff,2D}^{4}$. With that relation we can define the fraction of collapsed mass as:
\be
f_{\rm coll} = \rm min\left( 1, \dfrac{t_{\rm fb}}{t_{\rm ff, 2D}}\right)^4,
\ee
\noindent
which will be less than unity when the stellar feedback halts collapse towards faster than the two-dimensional collapse is completed.

Thus, the maximum mass of the collapsing region can be determined from the Toomre mass and the collapsed-mass fraction as:
\be
M_{\rm GMC, max} = M_{\rm T}\times f_{\rm coll}=\dfrac{4\pi^5G^2\Sigma_{\rm g}^3}{\kappa^4}\times  \rm min\left( 1, \dfrac{t_{\rm fb}}{t_{\rm ff, 2D}}\right)^4.
\label{eq:maxmassgmc}
\ee
\noindent
Inspection of equations~(\ref{eq:tff-2D}) and (\ref{eq:tff-g}) shows that feedback dominates in regions of low epicyclic frequency and low gas surface density, whereas free-fall dominates in regions of high epicyclic frequency and/or high gas surface density. We refer to these two regimes as feedback-limited and shear-limited, respectively.

\subsection{Mass of the most massive stellar cluster}

Given the mass of the most massive GMC, it is possible to derive the mass of the most massive stellar cluster. We assume that cluster complexes can continue to collapse within the collapsed region of the cloud even if the stellar feedback stops the collapse of the GMC. Following \citet{kruijssen14c}, we derive the maximum cluster mass by assuming some fraction of the GMC is converted into stars (i.e.~the star formation efficiency or SFE, $\epsilon$), of which some fraction is born in gravitationally bound stellar clusters (i.e.~the cluster formation efficiency or CFE, $\Gamma$ as defined by \citealt{bastian08}), resulting in:
\be
M_{\rm cl, max} = \epsilon \Gamma (\Sigma, \kappa, Q) M_{\rm GMC, max},
\label{eq:mass-cluster}
\ee
where we set $\epsilon = 0.1$ \citep[e.g.][]{lada03, oklopcic17}. This fiducial value is broadly consistent with the SFE of embedded clusters ($0.1$--$0.3$, \citealt{lada03}) and nearby molecular clouds ($0.03$--$0.06$, \citealt{evans99}). The total dynamic range of the SFE is there for a factor of three in either direction. The maximum cluster mass-scales in this paper cover an range of several orders of magnitude (see Section~\ref{sec:testmethod}). Assuming a constant SFE is therefore a minor effect relative to the other elements in our model (such as the steep dependences on the epicyclic frequency and the surface density). The CFE $\Gamma(\Sigma,\kappa,Q)$ is evaluated at $t=t_{\rm fb}$ using the model from \citet{kruijssen12d}. This model predicts the naturally bound fraction of star formation $f_{\rm bound}$ as well as the fraction thereof that survives tidal perturbations by nearby gas clouds on a star formation time-scale (`the cruel cradle effect’). Because we are interested in the most massive cluster that can possibly form, we adopt $\Gamma=f_{\rm bound}$ and ignore the cruel cradle effect. For the same reason, we consider the limiting case in which the cloud undergoes complete hierarchical merging and all bound stars end up in a single cluster.

\section{Exploring parameter space} \label{sec:results}

Using the model derived in the previous section, we now calculate the resulting maximum GMC and cluster mass-scales, and demonstrate in which part of the parameter space formed by $\Sigma_{\rm g}$, $\kappa$ and $Q$ these are limited by feedback rather than by shear. 

The model has been derived without the assumption of a flat rotation curve. In order to facilitate the comparison to observations, we now assume a flat rotation curve and change the epicyclic frequency $\kappa$ in equations ~(\ref{eq:tff-2D}), ~(\ref{eq:tfbgmc}), ~(\ref{eq:maxmassgmc}) and ~(\ref{eq:mass-cluster}), following:
\be
\kappa \equiv \sqrt{2} \dfrac{V}{R}\sqrt{1+\dfrac{{\rm d}\ln V}{{\rm d}\ln R}} \rightarrow \sqrt{2}\Omega,
\label{eq:kappa}
\ee
\noindent
where $V$ corresponds to the circular velocity, $R$ corresponds to the galactocentric radius and $\Omega$ is the angular velocity, i.e.~$\Omega=V/R$.

We start by setting a fiducial value of the stability parameter $Q = 1.5$ to study the predictions of our model as a function of the gas surface density and the angular velocity, and then we study the influence of the Toomre parameter on our predictions.

\subsection{Influence of the gas surface density and the angular velocity}

\begin{figure*}
	\includegraphics[width=\hsize]{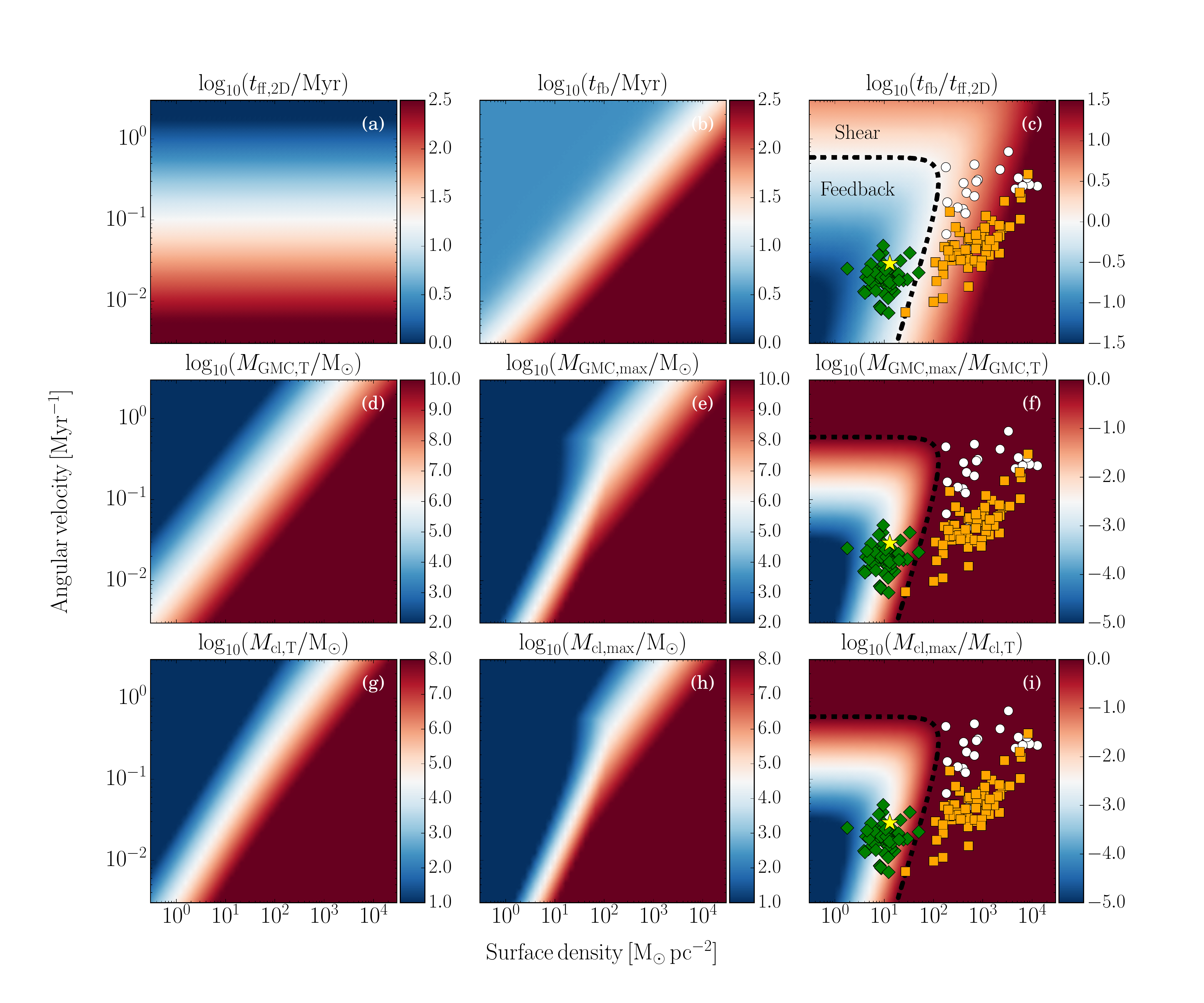}
	\caption{
		\label{fig:fig1-general}
		Predictions of our model assuming $Q = 1.5$. From left to right, the panels represent: (\textit{1st row}) the two-dimensional free-fall time-scale of a GMC $t_{\rm ff, 2D}$, the feedback time-scale $t_{\rm fb}$ and the ratio of time-scales $t_{\rm fb}/t_{\rm ff, 2D}$; (\textit{2nd row}) the shear-limited Toomre mass of the cloud $M_{\rm GMC,T}$, the hybrid maximum mass of the cloud $M_{\rm GMC, max}$ and the ratio of cloud mass-scales; (\textit{3rd row}) the shear-limited mass of the cluster $M_{\rm cl,T}$, the hybrid maximum mass of the cluster $M_{\rm cl, max}$ and the ratio of cluster mass-scales. We overplot observational data from four different environments. The values for the solar neighbourhood are represented with a yellow star. The green diamonds represent the spiral galaxies from Table 1 in \citet{kennicutt98}, whereas the circumnuclear starbursts of their Table 2 are represented by white dots. The orange squares correspond to the high-redshift galaxies from \citet{tacconi13}.}
\end{figure*}

The results for a fiducial value of $Q = 1.5$ are presented in Fig.~\ref{fig:fig1-general}. From left to right and top to bottom, the panels represent the two-dimensional free-fall time-scale of the GMC ($t_{\rm ff, 2D}$), the feedback time-scale ($t_{\rm fb}$), the ratio of time-scales, the shear-limited GMC mass ($M_{\rm GMC,T}$), the hybrid maximum GMC mass including feedback ($M_{\rm GMC,max}$), the ratio of both GMC mass-scales, the shear-limited cluster mass ($M_{\rm cl,T}$), the hybrid maximum cluster mass including feedback ($M_{\rm cl,max}$), and the ratio of both cluster mass scales.

For comparison, we also overplot the positions in parameter space of four observed galaxy samples: the solar neighbourhood, nearby spiral galaxies, circumnuclear starbursts and high-redshift galaxies. For the solar neighbourhood we use the observed values $\Sigma_{\rm g, MW} \simeq 13\,\msun\pc^{-2}$ (fig. 7 from \citealt{kennicutt12}) and $\Omega_{\rm MW}\simeq0.029\,\myr^{-1}$ (Section 6.4.2 in \citealt{blandhawthorn16}). The nearby spiral galaxies are listed in Table 1 of \citet{kennicutt98}, whereas the circumnuclear starburst galaxies are listed in their Table 2. Lastly, the high-redshift galaxies are listed in Table 2 of \citet{tacconi13}. For these, we determine the angular velocity from the listed circular velocity and the optical half-light radius and we use their measurement of the mean molecular gas surface density within the half-mass radius.

As expected from equation~(\ref{eq:tff-2D}), in panel (a) it is found that the two-dimensional free-fall time of the cloud does not depend on the gas surface density, but it is inversely proportional to the angular velocity. This dependency is especially relevant in environments with low angular velocity, as they yield very large collapsing time-scales. By contrast, the feedback time-scale depends on both parameters, as seen in panel (b). The feedback time-scale reaches its lower limit, $t_{\rm fb} = t_{\rm sn}$, in environments with high angular velocity and low gas surface density. 

The contrast in the behaviour between the two time-scales is best visualised in panel (c), where we show the ratio of the feedback time over the two-dimensional free-fall time-scale of the shear-limited area. The dashed line marks the separation between the shear-limited ($t_{\rm ff,2D} < t_{\rm fb}$) and the feedback-limited ($t_{\rm ff,2D} > t_{\rm fb}$) regimes. It can be seen that, for low angular velocity and low gas surface density, the feedback time-scale can be more than an order of magnitude less than the free-fall time-scale. This directly implies that shear cannot be setting the maximum mass-scales of GMCs and clusters for an important part of parameter space.

For a fixed value of the angular velocity, the GMC mass-scales (panels (d) and (e)) increase with increasing gas surface density, whereas they decrease for increasing angular velocity and a fixed gas surface density. The hybrid GMC mass behaves like the shear-limited Toomre mass except for low angular velocity and low gas surface density, where it has lower values due to the short feedback time-scales.

The ratio between these two GMC masses is shown in panel (f) and, as expected, the parameter space affected matches the area where the feedback time is shorter than the free-fall time. The region of the parameter space affected by our model is occupied by most spiral galaxies in the local Universe (here showing the sample from \citealt{kennicutt98}, including the solar neighbourhood). Higher-density systems, such as the starburst galactic nuclei from \citet{kennicutt98} and the high-redshift galaxies from \citet{tacconi13} are located in the shear-limited regime. This indicates that the higher GMC (or clump) masses observed in high-redshift and starburst galaxies are due to them being limited by shear instead of by stellar feedback.

Turning to maximum cluster masses in panels (g)-(i), we see that the parameter space affected by our model is still in agreement with the time-scale ratios in panel (c). Compared to the GMC mass-scales, the cluster mass-scales shown in panels (g) and (h) have an overall similar behaviour due to the dependence of the CFE on $\Sigma_{\rm g}$ and $\Omega$. The differences are due to the dependence of the CFE on $\Sigma_{\rm g}$ and $\Omega$.

\subsection{Influence of the stability parameter Q}

We now explore the influence of the Toomre parameter in our model. We use three values of $Q = 0.5, 1.5$ and $3$ and we consider the ratio of time-scales, the shear-feedback hybrid GMC mass, the ratio of GMC mass-scales, the shear-feedback hybrid cluster mass, and the ratio of cluster mass-scales shown in panels (c), (e), (f), (h) and (i) of Fig.~\ref{fig:fig1-general}, respectively. 

\begin{figure*}
	\centering
	\includegraphics[width=\hsize]{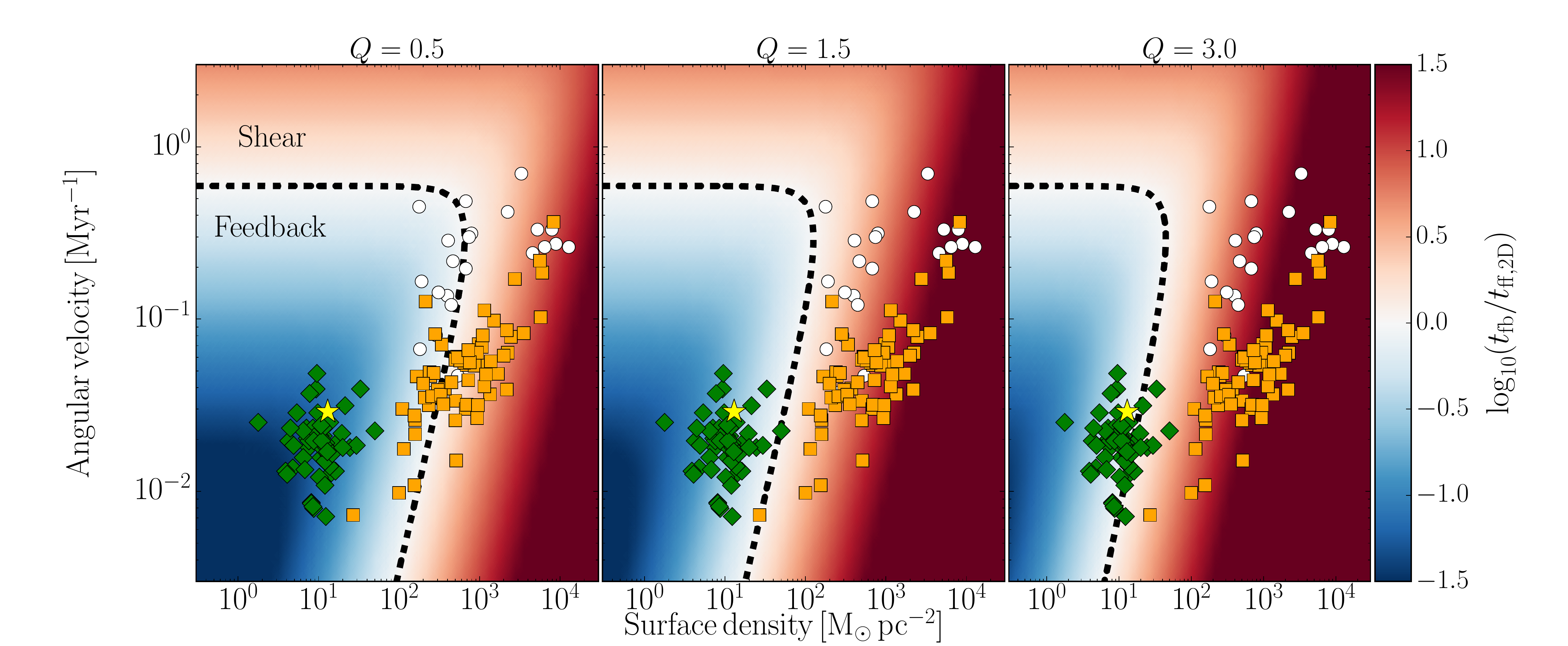}
	\caption{
		\label{fig:fig1-ratioT}
		Influence of the stability parameter $Q$ on the ratio of the feedback and free-fall time-scales $t_{\rm fb}/t_{\rm ff,2D}$: (\textit{left}) $Q = 0.5$, (\textit{middle}) $Q = 1.5$ and (\textit{right}) $Q = 3$. The dashed line indicates the change of regime from being feedback-limited ($t_{\rm ff,2D} > t_{\rm fb}$) in the bottom-left to being shear-limited ($t_{\rm ff,2D} < t_{\rm fb}$) in the top and/or right. We overplot observational data from four galactic environments as in Fig.~\ref{fig:fig1-general}.
	}
\end{figure*}

In Fig.~\ref{fig:fig1-ratioT}, we present the ratio of time-scales for the three values of the stability parameter. The region of the parameter space affected by our addition of feedback as a limiting mechanism increases as we move towards lower values of $Q$ (from right to left panels). The resulting reduction of the maximum mass-scales is caused by an increase of the gas surface density needed to go from the shear-limited to the feedback-limited regime, whereas the limit on the angular velocity does not change. This result indicates that environments with a low value of $Q$, which are less shear stable, are more likely to be limited by feedback rather than by shear.

\begin{figure*}
	\centering
	\includegraphics[width=\hsize]{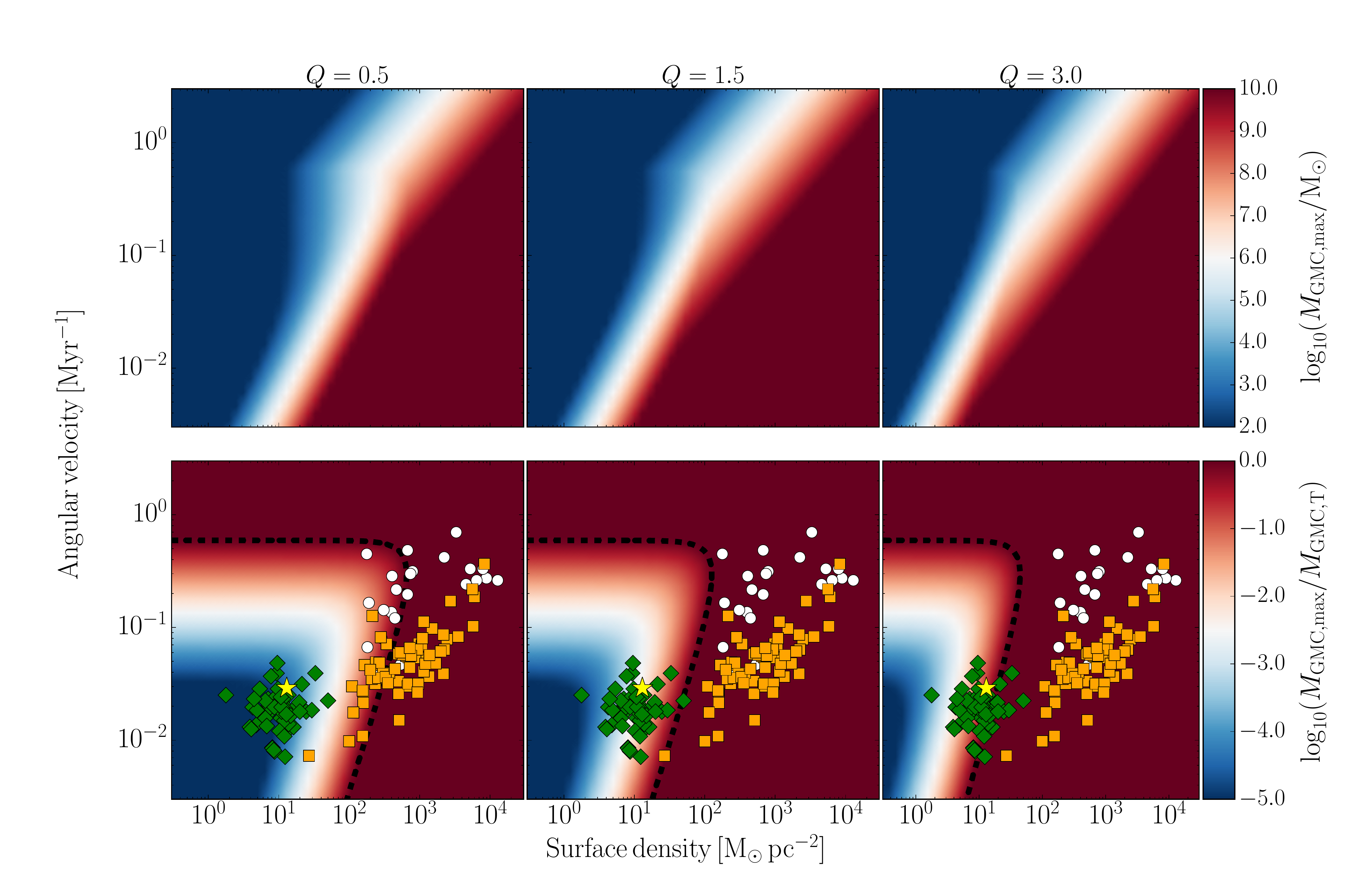}
	\caption{
		\label{fig:fig1-ratioMgmc}
		Influence of the stability parameter $Q$ on (\textit{top}) the maximum shear-feedback hybrid GMC mass and (\textit{bottom}) the ratio of GMC mass-scales: (\textit{left}) $Q = 0.5$, (\textit{middle}) $Q = 1.5$ and (\textit{right}) $Q = 3$. The dashed line indicates the change of regime from being feedback-limited ($t_{\rm ff,2D} > t_{\rm fb}$) in the bottom-left to being shear-limited ($t_{\rm ff,2D} < t_{\rm fb}$) in the top and/or right. We overplot observational data from four galactic environments as in Fig.~\ref{fig:fig1-general}.
	}
\end{figure*}

\begin{figure*}
	\centering
	\includegraphics[width=\hsize]{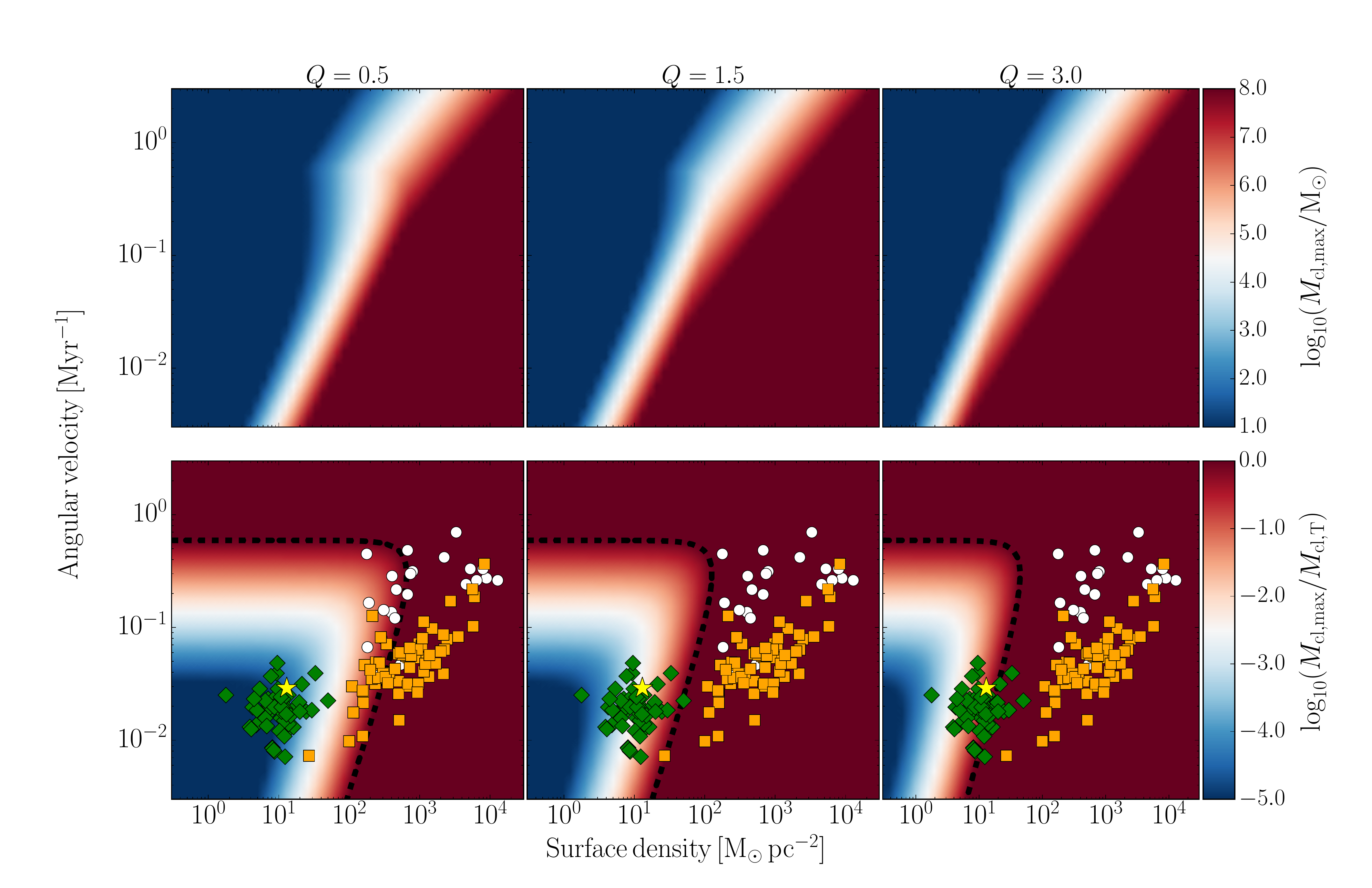}
	\caption{
		\label{fig:fig1-ratioMcl}
		Influence of the stability parameter $Q$ on (\textit{top}) the maximum shear-feedback hybrid cluster mass and (\textit{bottom}) the ratio of cluster mass-scales: (\textit{left}) $Q = 0.5$, (\textit{middle}) $Q = 1.5$ and (\textit{right}) $Q = 3$. The dashed line indicates the change of regime from being feedback-limited ($t_{\rm ff,2D} > t_{\rm fb}$) in the bottom-left to being shear-limited ($t_{\rm ff,2D} < t_{\rm fb}$) in the top and/or right. We overplot observational data from four galactic environments as in Fig.~\ref{fig:fig1-general}.
	}
\end{figure*}

The variation of $Q$ has a similar effect on the GMC and cluster mass-scales, shown in the top panels of Figs.~\ref{fig:fig1-ratioMgmc} and~\ref{fig:fig1-ratioMcl}, respectively. The white region, corresponding to $M_{\rm GMC, max} \sim 10^6\,\rm M_{\odot}$ and $M_{\rm cl, max} \sim 10^{4.5}\,\rm M_{\odot}$, shifts towards lower gas surface densities as we move towards larger values of $Q$ (from left to right), thus indicating that the more shear-stable environments will have larger cloud masses compared to the unstable environments. 

\citet{leroy08} show in their fig.~11 that nearby galaxies have an observed Toomre parameter $Q\sim3$, which would place them in the region of the parameter space affected by our introduction of the feedback mechanism. By contrast, and despite the low value of the stability parameter of $Q\sim1$ reported in fig.~24 in \citet{genzel14}, the high-redshift galaxies lie in the region of the parameter space that is only limited by shear. It seems that as galaxes evolve and becomes less gas-rich, they move in the parameter space from the shear-limited towards the feedback-limited region.

\subsection{Limitations of the model}

In Section~\ref{sec:model} we derive our model by considering a differentially rotating disc in hydrostatic equilibrium. This assumption implies the presence of some degree of shear and allows us to describe the properties of the ISM just using three variables ($\Sigma$, $\kappa$ and $Q$). These simple assumptions yield some caveats that need to be accounted for when comparing the predictions of our hybrid model with the observed mass-scales of real galaxies.
	
The first caveat is due to our model predicting the maximum mass-scales based on the present-day properties of the gas surface density. Therefore, the predictions of the hybrid model should be compared to a cluster sample with an age range small enough such that the gas properties have not undergone significant change. Otherwise, the input gas conditions in our model will not agree with the gas conditions at birth of these clusters. A natural time-scale for changes in the gas conditions is the orbital time-scale $\Omega^{-1}$, which is of the order $50\,\myr$ in nearby galaxies \citep{leroy08}. The second caveat comes from the fact that our hybrid shear-feedback model does not consider systematic structural morphological features that may exist in the disc, such as bars, rings or spiral arms. Such structures may induce the formation of more massive clouds than the ones predicted by our model, as the external compression from the structure would yield larger densities and thus larger mass-scales that would not have been reached if shear is considered. Finally, we also need to consider the effect of dynamical friction on the most massive clusters when we apply our model to star-forming high-redshift environments. According to the cluster formation model described in \citet{kruijssen15b}, stellar clusters may survive over a Hubble time and become the present-day GCs if they migrate towards the host galaxy halo (e.g.~by hierarchical galaxy growth) before they are destroyed by impulsive shocks or dynamical friction into the centre. For the most massive clusters, the dynamical friction time-scale may be too short for the clusters to survive till migration, which would cause them to spiral in and contribute to the growth of a bulge. Given a galaxy stellar mass, it is possible to determine what is the maximum mass of clusters that will survive dynamical friction, i.e.~the maximum mass that may be observed in the local Universe. We will have to correct the predictions of our model if they are larger than these dynamically limited mass-scales (see Section~\ref{subsec:pred-zc406690}).

\section{Comparison to observations of clouds and clusters}\label{sec:testmethod}

We now apply our shear-feedback hybrid model to observational data and compare the results with the observed GMCs and cluster masses, from local galaxies out to the clumpy, star-forming systems observed at high redshift. We choose four galactic environments as probes to test the model: the Milky Way, the disc galaxy M31, the grand-design spiral galaxy M83 and the high-redshift galaxy ZC406690. These four galaxies represent the subset of systems for which the most comprehensive observational data are available to carry out a detailed comparison and, at the same time, they have a sufficiently wide variety of properties to cover the galaxy population of interest, including different galactic stellar masses ($\sim 10^{10}$--$10^{11}\,\msun$), star formation rate surface densities ($\sim 10^{-3}$--$10^{1}\,\msun\,\kpc^{-2}\,\yr^{-1}$), substructure (spiral arms, clumpy and ring-like morphologies), and gas fraction ($\sim10$--$70$ per cent). 

\subsection{Milky Way} \label{sec:predictions-mw}

We carry out a first test of our model by simultaneously comparing it to the most massive GMCs and stellar clusters in our Galaxy. We apply the model to two different regions of the Milky Way: the Central Molecular Zone (corresponding to $R<0.2\,\rm kpc$, CMZ hereafter) and the disc (corresponding to $R\in[3.5,10]\,\rm kpc$). For the CMZ, we use a gas surface density of $\Sigma_{\rm g}\sim 10^3\,\rm M_{\odot} pc^{-2}$ and a one-dimensional velocity dispersion of $\sigma = 5\,\rm km\, s^{-1}$ (as in \citealt{henshaw16}) and we determine the circular velocity from the relation between the enclosed mass and the radius shown in fig.~A1 in \citet{kruijssen15}. For the disc of the Milky Way, we use the red rotation curve shown in fig.~16 in \citet{blandhawthorn16} for a disc with a radial scale-length of $R_{\rm d} = 2.6\,\rm kpc$. To ensure the continuity of the derivatives, we carry out a b-spline fit the rotation curve before calculating the epicyclic frequency as in equation~\ref{eq:kappa}. We determine the total gas surface density from the addition of the atomic and molecular gas surface densities shown in fig.~7 in \citet{kennicutt12}\footnote{We repeat the calculation using the gas surface density profile in fig.~9 of \citealt{mivilledeschenes17} and we obtain the same mass-scale predictions to within the errorbars.} and we assume a fiducial constant one-dimensional velocity dispersion of $\sigma = 10\,\rm km\, s^{-1}$ (cf.~\citealt{heiles03}). This allows us to determine the stability parameter $Q$ as a function of the galactocentric radius as in equation~\ref{eq:toomreQ}.

\begin{figure}
	\includegraphics[width=\hsize]{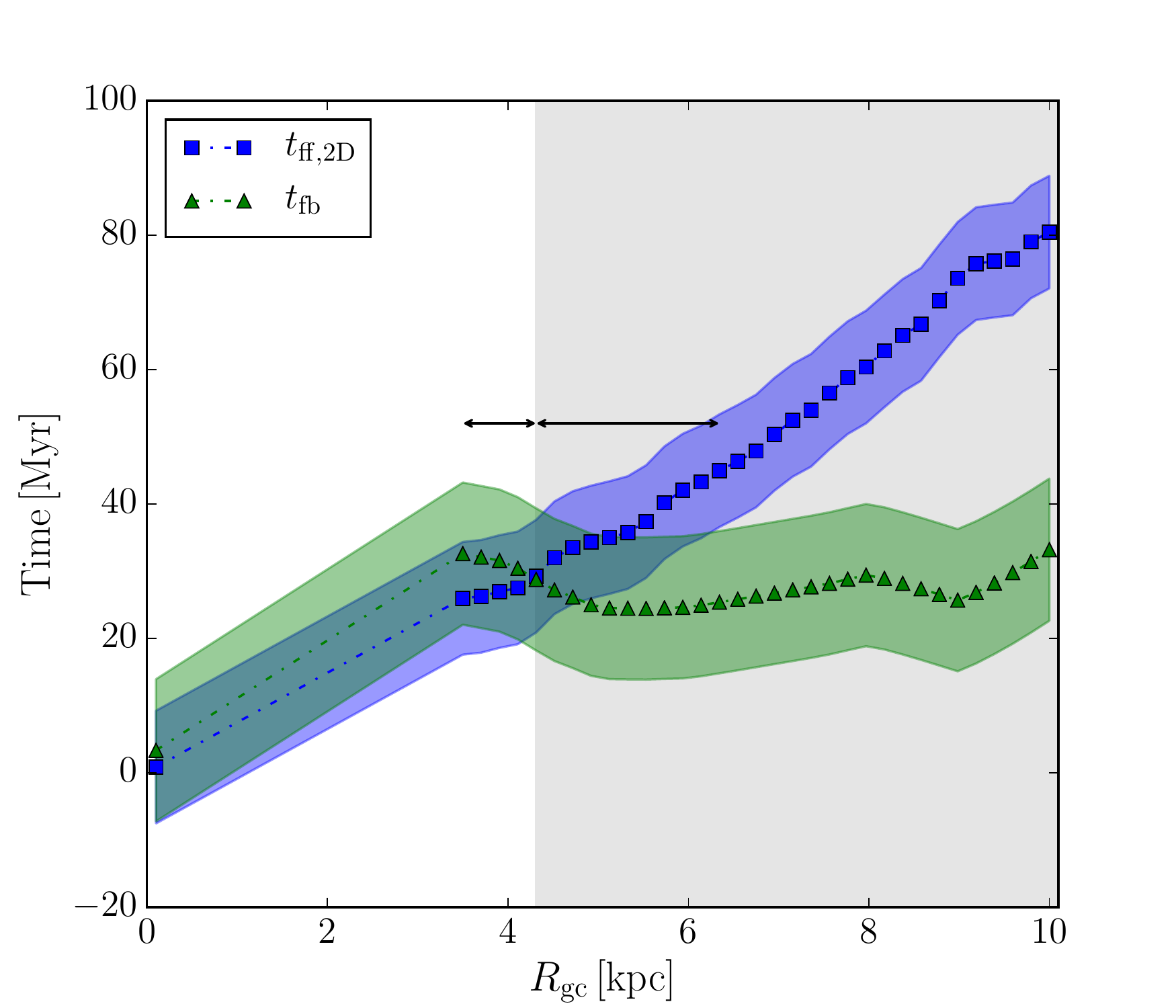}
	\caption{
		\label{fig:fig2-time-scales}
		Predicted free-fall and feedback time-scales for GMCs in the Milky Way as a function of the galactocentric radius. The colour-shaded area indicates the uncertainties associated with the characteristic time-scales determined with $10^6$ Monte-Carlo runs assuming typical uncertainties for the epicyclic frequency and the gas surface density of $10$ per cent and $30$ per cent, respectively. The grey-shaded area corresponds to the feedback-limited regime, $t_{\rm fb} < t_{\rm ff, 2D}$, located at $R\geq4.3^{+2.0}_{-0.8}\,\kpc$. The lower errorbar is determined from the inner boundary of the data, whereas the higher one comes from the change of regime of the colour-shaded areas. Both of them are represented with black arrows.
	}
\end{figure}

With the values obtained, we evaluate the model of Section~\ref{sec:model} at each galactocentric radius. We determine the free-fall and feedback time-scales, as well as the maximum GMC and cluster mass-scales. We show a comparison of the two-dimensional free-fall time and the feedback time in Fig.~\ref{fig:fig2-time-scales}, where we determine the uncertainties associated to the characteristic time-scales performing $10^6$ Monte-Carlo runs assuming typical uncertainties of 10 per cent and 30 per cent for the epicyclic frequency and the gas surface density, respectively. We shade the area in Fig.~\ref{fig:fig2-time-scales} where the feedback time-scale is smaller than the collapsing free-fall time, i.e.~when the cloud masses will be feedback-limited and less massive than the Toomre mass. This happens at galactocentric radii  $R \geq 4.3^{+2.0}_{-0.8}\,\kpc$ due to a gradual change of the time-scales with the environment, not due to any particular (morphological) feature such as the tip of the bar. For smaller galactocentric radii, the two-dimensional free-fall time is shorter than the feedback time and the maximum cloud mass is equal to the Toomre mass.

\begin{figure}
	\includegraphics[width=\hsize]{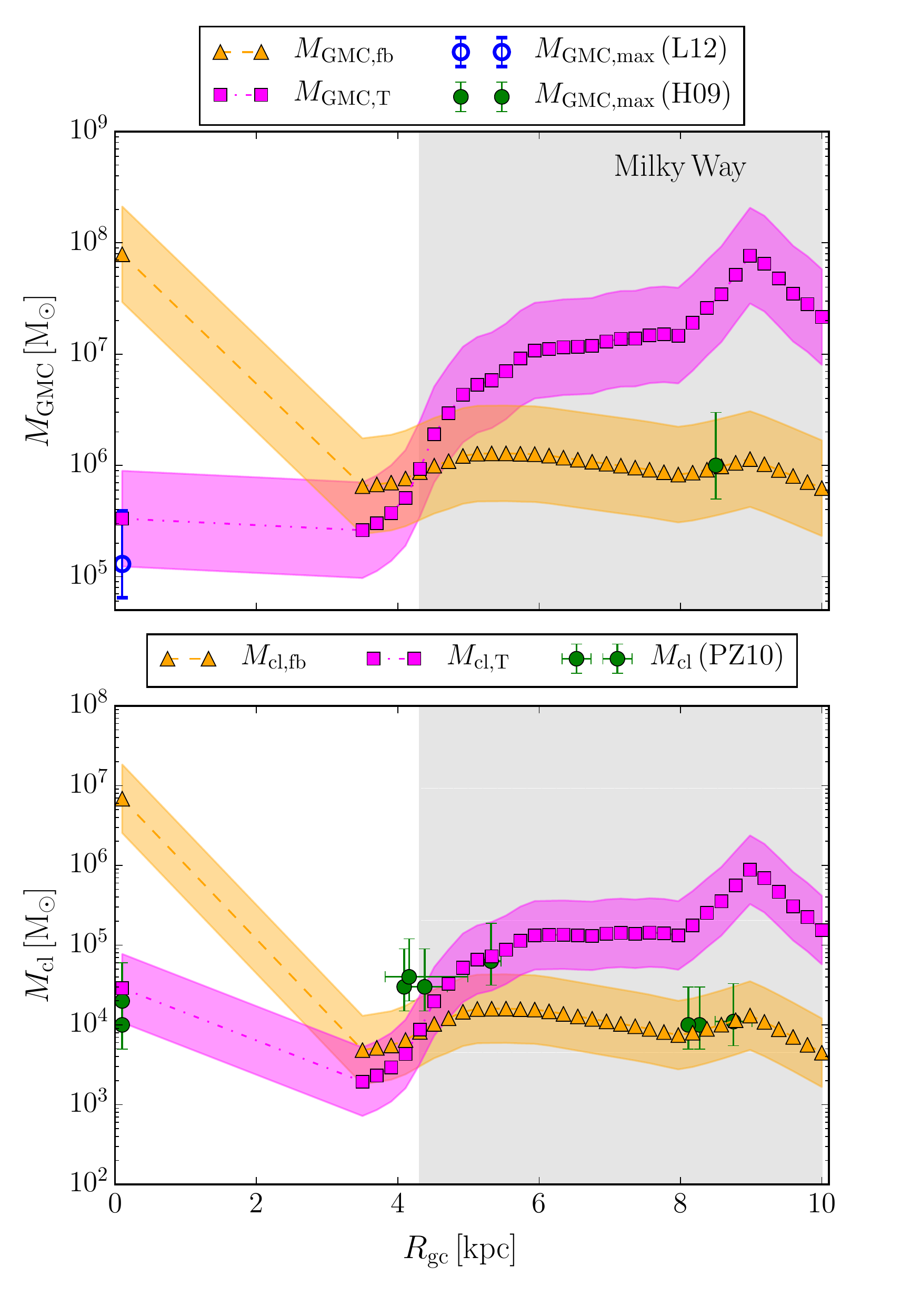}
	\caption{
		\label{fig:fig3-mwg}
		Result of applying our shear-feedback hybrid model to the Milky Way: (\textit{top}) maximum GMC mass and (\textit{bottom}) maximum cluster mass as a function of the galactocentric radius. The magenta squares correspond to the shear-limited mass-scales and the orange triangles correspond to the feedback-limited mass-scales. The observed maximum GMC and cluster masses are represented by the green and blue dots with error bars. The grey-shaded area corresponds to the feedback-limited regime: $t_{\rm fb} < t_{\rm ff, 2D}$. The colour-shaded areas indicate a fiducial uncertainty range, assuming that the rotation curve and the gas surface density profile are known to an accuracy of $\sim0.04$~dex and $\sim0.13$~dex, respectively.}
\end{figure}

The panels in Fig.~\ref{fig:fig3-mwg} show the maximum cloud mass and the maximum cluster mass, respectively. For the sake of a better visualization, we separately show the maximum mass set by each mechanism (i.e. shear and feedback). We remind the reader that the lowest of these curves sets the maximum mass-scale. With typical uncertainties of $10$ per cent in the epicyclic frequency and $30$ per cent in the gas surface density, the uncertainty associated with our predictions is $\sim0.4\,\rm dex$. In the CMZ, the clouds are shear-limited and our prediction agrees well with the observed cloud mass of $\sim10^5~\msun$ in \citet{longmore12}. For $R \gtrsim 4.3\,\kpc$, the mass-scales become feedback-limited, as indicated by the shaded area. In that regime, the feedback-limited GMC mass remains approximately constant at $M_{\rm GMC, max}\sim10^6\,\rm M_{\odot}$, in agreement with the most massive clouds for the solar neighbourhood reported in fig.~3 in \citet{heyer09}\footnote{These authors first determine the $\textsuperscript{13}{\rm C}$ and $\textsuperscript{12}{\rm C}$ column densities from the $\textsuperscript{13}{\rm CO}$($1$--$0$) emission and using a radially-dependent conversion factor of $\textsuperscript{12}{\rm C}/\textsuperscript{13}{\rm C}$ taken from \citet{milam05}, respectively. They then determine the $H_2$ column density by assuming a constant H$_2$/$\textsuperscript{12}{\rm CO}$ abundance ratio of $1.1\times10^4$ (\citealt{frerking82}).}. The predicted shear-limited cloud and cluster masses increase at large galactocentric radius (for $R > 4.3\,\rm kpc$) due to the steady drop of the epicyclic frequency. The resulting increase of the two-dimensional free-fall time implies that feedback becomes the mechanism responsible for setting the approximately constant GMC mass as a function of radius.

In the bottom panel, we show the predicted maximum cluster masses by our model. We overplot the masses of the observed clusters Arches, Quintuplet, RSGC01, RSGC02, RSGC03, Westerlund 1, Westerlund 2, Trumpler 14 and NGC 3603 reported in Table~2 in \citet{portegieszwart10}. The vertical error bars correspond to an uncertainty of $\pm 0.3\,\rm dex$, whereas the error bars on the galactocentric radius have been propagated from the distance uncertainties in the original papers referenced by \citet{portegieszwart10}. Except for the clusters located at the end of the bar ($R\simeq 4 \,\rm kpc$), the predicted cluster masses for the CMZ and the solar neighbourhood are in agreement with the observed cluster masses. A possible reason why the cluster masses at the end of the bar would be elevated is that the bar may be sweeping up the material, thus producing an environment in which external compression aids mass accumulation towards mass-scales that otherwise would not be achieved. The data point at $R\approx5.5\,\kpc$ corresponds to Westerlund 1, which resides in the Scutum arm that connects to the bar. It lies just 10 Myr downstream from the bar, which means that it likely formed through a similar compression event like we suggest for the RSG clusters (located at $R\approx4\,\kpc$).

\subsection{M31}

We now test our model with the disc galaxy M31, the most massive galaxy in the Local Group. We apply the model in the radial range $R\in[8,15]\,\kpc$ in order to have measurements for all input variables. This radial range contains the $10$-kpc ring of star formation, i.e.~the region where cluster formation is more likely to occur (\citealt{vansevicius09}, \citealt{caldwell09} and references therein). We use the turbulent velocity dispersion from \citet{braun09} (fig. 19, panel f) and the rotation curve described in \citet{corbelli10} (fig.~7 and Table~1), both of them obtained with the Westerbork Synthesis Radio Telescope (WRST) HI survey of M31 described in \citealt{braun09}). We determine the gas surface density from the addition of the atomic and molecular profiles reported in \citet{schruba17}, of which the data are described by \citet{leroy16}, and determined from observations with the WRST and the IRAM 30m telescope, respectively.

\begin{figure}
	\includegraphics[width=\hsize]{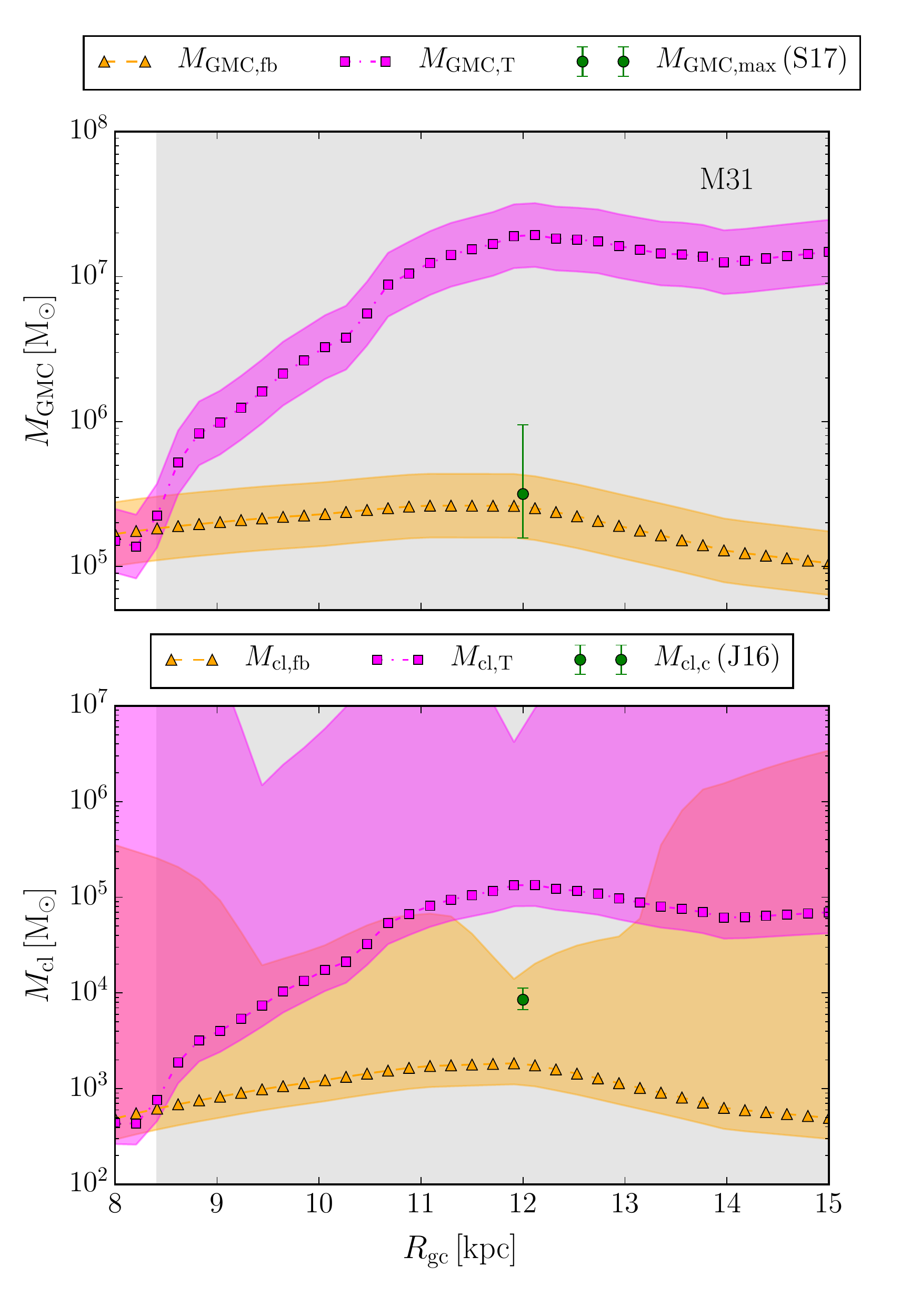}
	\caption{
		\label{fig:fig3-m31}
		Result of applying our shear-feedback hybrid model to the disc galaxy M31: (\textit{top}) maximum GMC mass and (\textit{bottom}) maximum cluster mass as a function of the galactocentric radius. The magenta squares correspond to the shear-limited mass-scales and the orange triangles correspond to the feedback-limited mass-scales. The observed maximum GMC and cluster masses are represented by the green and blue dots with error bars. The grey-shaded area corresponds to the feedback-limited regime: $t_{\rm fb} < t_{\rm ff, 2D}$. The colour-shaded areas indicate a fiducial uncertainty range, assuming that the rotation curve and the gas surface density profile are known to an accuracy of $\sim0.04$~dex. It also accounts for changes in the gas surface density over the past $300~\myr$, corresponding to the age range for which the cluster masses have been measured.}
\end{figure}

As previously done for the Milky Way, we do not assume a flat rotation curve, but instead we determine the epicyclic frequency as in equation~(\ref{eq:kappa}) and carrying out a b-spline fit to the rotation curve. We then determine the stability parameter $Q$ at each galactocentric radius as in equation~(\ref{eq:toomreQ}). With these values, we apply the model described in Section~\ref{sec:model} and show the predicted cloud and cluster mass-scales in Fig.~\ref{fig:fig3-m31}. 

In the top panel we represent the predicted shear and feedback-limited mass-scales for the GMCs. With typical uncertainties of $10$ per cent for both the gas surface density\footnote{This uncertainty is lower than for the Milky Way, where we assumed $30$ per cent, because M31 is atomic gas-dominated.} and the epicyclic frequency, we determine the uncertainties associated with our predictions to be $\sim0.22\,\rm dex$. As indicated by the shaded area, the maximum mass-scales are feedback-limited for galactocentric radii $R\geq8.4\,\kpc$, whereas they become shear-limited at smaller radii. There is a good agreement between our predictions and the observed maximum GMC mass of $M_{\rm GMC,max} \sim 10^{5.5}\,\msun$ reported in \citet{schruba17}.\footnote{The cloud mass is derived from CO data using the standard Galactic CO-to-H$_2$ conversion factor of $X_{\rm CO} = 2\times 10^{20}\,\rm cm^{-2}\,(K\,\kms)^{-1}$ (\citealt{bolatto13}).}

The maximum cluster mass-scales are shown in the bottom panel of Fig.~\ref{fig:fig3-m31}. The symbols correspond to the prediction of our model using the current gas condition, whereas the colour-shaded area reflects a fiducial uncertainty range of $\sim0.22\,\rm dex$, as well as changes in the gas surface density over the past $300~\myr$, corresponding to the age range for which the cluster masses in \citet{johnson17} have been measured. These authors determine the observed truncation mass by fitting an exponentially truncated power-law with index -2 (i.e.~a \citealt{schechter76} function) to the mass function of a sample of 840 young clusters of ages between $10$--$300~\myr$. Given this large age range, it may not be surprising that the present-day predictions of our model do not reproduce the observed maximum cluster mass-scale. The $10$-kpc ring is a long-lived structure with a bursty star formation history \citep{lewis15}, which means that the currently-observed gas properties may not be representative for the conditions under which the most massive clusters in the ring formed. Indeed, figs. 5 and 6 of \citet{lewis15} show that, over the past $300\,\myr$, the star formation rate surface density of the $10$-kpc ring experienced several peaks of up to a factor of $4$ higher than it is at present. This increase would correspond to a similar increase of the CFE and, hence, of the maximum cluster mass.

In order to account for the past variations of the gas surface density, we assume a linear relation between the SFR density and the gas surface density,  $\Sigma_{\rm g}\propto\Sigma_{\rm SFR}$ (\citealt{bigiel08}, \citealt{leroy08}), and use fig.~6 in \citet{lewis15} to derive a radially dependent correction factor, $f(R)\equiv \max{[\Sigma_{\rm SFR}(R,0\leq \tau/{\rm Myr}<316)]}/\Sigma_{\rm SFR}(R,0\leq \tau/{\rm Myr}<25)$. This factor converts the SFR over the most recent $25\,\myr$, and hence, the present gas surface density, to the maximum over the past $316\,\myr$, as that corresponds to the true conditions when the clusters were formed. We apply our model to the past-corrected gas surface density profile and those predictions are included in the colour-shaded area. We find that the predictions based on the present-day SFR underpredict the truncation mass reported in \citet{johnson17} which is $M_{\rm cl, c} = 8.5^{+2.8}_{-1.8}\times10^3\,\msun$ by a factor of $4$, but when accounting for the past range of gas surface densities, we find that our prediction agrees with the observed mass.
	
The above considerations show that at certain times in the history of the $10$-kpc ring, the maximum cluster mass-scales predicted by our model would have been consistent with the observed truncation mass from \citet{johnson17}. We can test further whether this is indeed the reason for the discrepancy by considering the ages of the most massive clusters in M31. \citet{fouesneau14} present a subset of the cluster sample used by \citet{johnson17} and show that the most massive clusters in the $0$--$300\,\myr$ age range have ages exceeding $100\,\myr$. This corresponds to multiple dynamical times at the radius of the $10$-kpc ring and matches a burst that occurred $100$--$150\,\myr$ ago. We thus conclude that the discrepancy between model and observation is likely due to changes in environmental conditions since the observed clusters formed.

\subsection{M83} \label{sec:predictions-m83}

For the third test of our model, we use the nearby grand-design spiral galaxy M83. In order to apply our model, we use the observational data reported in fig.~6 in \citet{freeman17}: the total gas surface density curve from the top panel, the rotation curve from the middle panel obtained from the 21-cm line in \citet{walter08} and the velocity dispersion inferred from the Atacama Large Millimeter/submillimeter Array (ALMA) CO and THINGS 21-cm data from the bottom panel. 

We compare the predictions of our model with the observed maximum masses of two observational samples. For the GMC sample, we use the cloud catalogue of \citet{freeman17}, who present data obtained with ALMA, whereas for the clusters, we use the sample from \citet{adamo15b}, based on Hubble Space Telescope (HST) data.

\begin{figure}
	\includegraphics[width=\hsize]{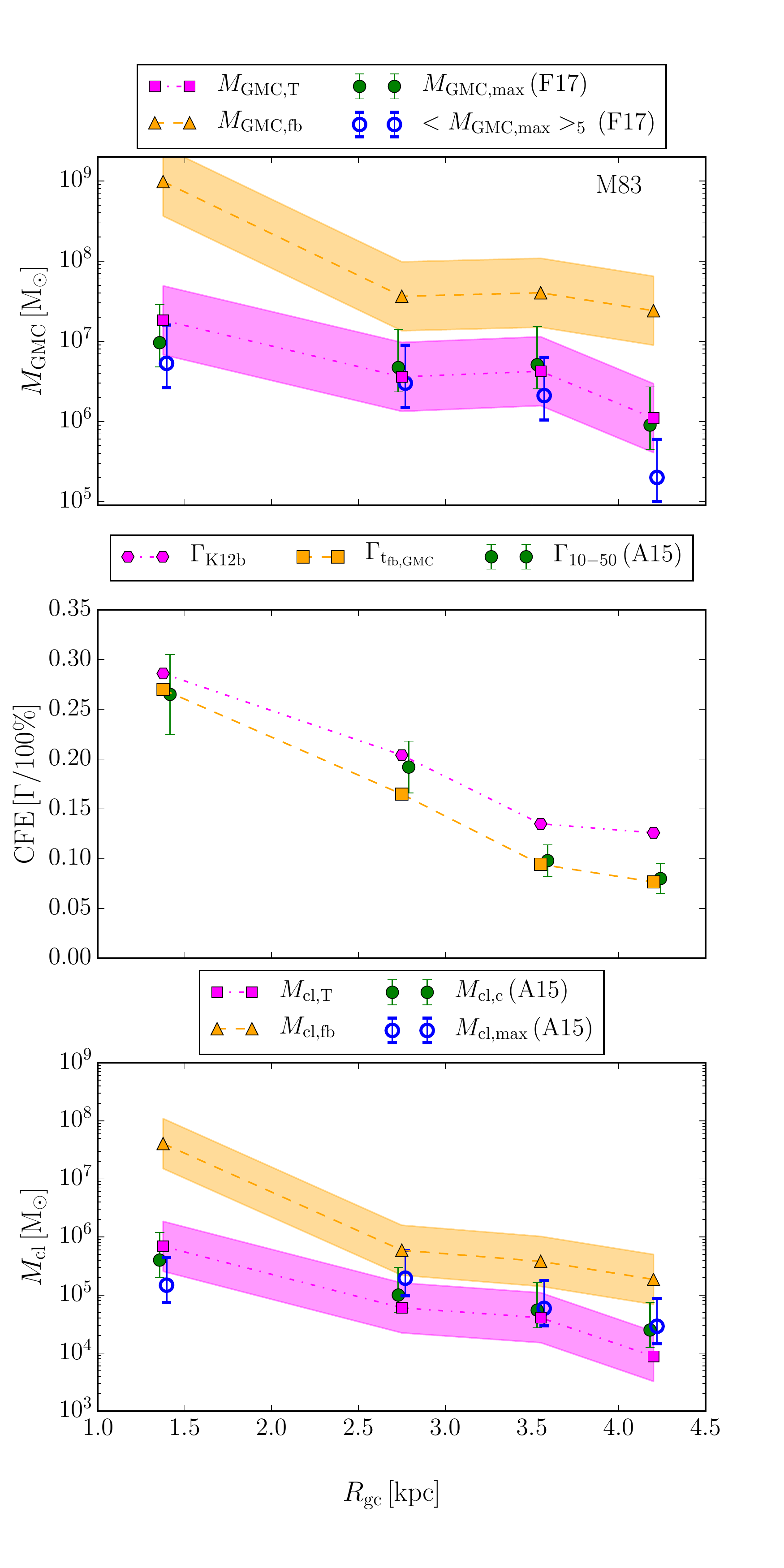}
	\caption{
		\label{fig:fig3-m83}
		Result of applying our shear-feedback hybrid model to M83: (\textit{top}) maximum GMC mass and (\textit{bottom}) maximum cluster mass as a function of the galactocentric radius. The magenta squares correspond to the shear-limited mass-scales and the orange triangles correspond to the feedback-limited mass-scales. The observed maximum mass-scales and CFE are represented by the green and blue dots with error bars. The colour-shaded areas indicate a fiducial uncertainty range, assuming that the rotation curve and the gas surface density profile are known to an accuracy of $\sim0.04$~dex and $\sim0.13$~dex, respectively. For more details see the text.}
\end{figure}

As for the previous tests, we carry out a b-spline fit to the rotation curve to determine the epicyclic frequency $\kappa$. We determine the stability parameter $Q$ as in equation~(\ref{eq:toomreQ}) and we apply the model from Section~\ref{sec:model} at each galactocentric radius. We determine the maximum cloud mass, the CFE, and the maximum cluster mass. In order to compare with the observed masses, we calculate the mass-weighted average of the predicted quantities for each radial bin described in \citet{adamo15b} and show the result in Fig.~\ref{fig:fig3-m83}. With typical uncertainties of $10$ per cent in the epicyclic frequency and $30$ per cent in the gas surface density, the uncertainty associated with our predictions is $\sim0.4\,\rm dex$.

In the top panel in Fig.~\ref{fig:fig3-m83}, we show the GMC mass-scales. We plot the mean mass for the five most massive GMCs $<M_{\rm GMC, max}>_5$ and the most massive GMC $M_{\rm GMC, max}$ reported in \citet{freeman17}\footnote{The cloud masses are derived from CO data using the standard Galactic CO-to-H$_2$ conversion factor of $X_{\rm CO} = 2\times 10^{20}\,\rm cm^{-2}\,(K\,\kms)^{-1}$ (\citealt{bolatto13}).}, as well as our predicted shear and feedback-limited mass-scales for the clouds. Given that the region studied is shear-limited, the predicted maximum mass by our model corresponds to the Toomre mass and they agree with the observed cloud masses, at each radial bin.

\citet{adamo15b} determine the CFE of two groups of clusters; $\Gamma_{1-10}$ for the clusters of ages between $1$--$10\,\myr$ and $\Gamma_{10-50}$ for the clusters of ages between $10$--$50\,\myr$. Given that for the age bin $1$--$10\,\myr$ it is non-trivial to distinguish between bound clusters and unbound associations, we decide to use $\Gamma_{10-50}$ to represent the CFE of the cluster population in M83.

We show the observed and predicted CFEs in the middle panel in Fig.~\ref{fig:fig3-m83}. Specifically, we show the observed CFE $\Gamma_{10-50}$ described in \citet{adamo15b} (the error bars correspond to the values given in their paper), the predicted CFE using the model described in \citet{kruijssen12d} with a time of observing the cluster population of $t = 10\,\myr$, and the predicted CFE using the same model but at a different time $t = t_{\rm fb}$. Our prediction agrees well with the observational $\Gamma_{10-50}$ in all the radial bins. To first order, this confirms the agreement found by \citet{adamo15b}, who carried out a similar comparison using lower-resolution input data for the model. To second order, we see that evaluating the CFE predicted by \citet{kruijssen12d} at the feedback time yields somewhat better agreement with the observations than using the fiducial evaluation time of $10\,\myr$. However, it is not immediately clear how significant this improvement is, given the intrinsic $\sim 0.2$~dex uncertainty in the model. Only at the largest radius does the difference between both models exceed this margin.

In the bottom panel of Fig.~\ref{fig:fig3-m83} we present the observed and predicted cluster masses. We show the maximum cluster mass $M_{\rm cl, max}$ and the characteristic cluster mass $M_{\rm cl, c}$ obtained when fitting an exponentially truncated power law with index $-2$ (i.e.~a \citealt{schechter76} function) presented in \citet{adamo15b}, as well as our predicted shear-limited (using $M_{\rm GMC} = M_{\rm GMC, T}$) and feedback-limited cluster masses. As in the cloud mass-scales, we predict the masses are shear-limited over the entire radial range and they closely agree with the observed masses. 

The strong bar in M83 only appears to slightly affect the observed cluster masses. From the maps in \citet{adamo15b}, the bar ends at $R\sim2.3\,\kpc$, which corresponds to the end of their first bin. Any influence of the bar should manifest itself in the first two radial bins. As opposed to the comparison done in Section~\ref{sec:predictions-mw} for the Milky Way, where we use individual masses of observed clusters, here we compare the predictions of our model with two mass-scales: the characteristic mass derived from a Schechter function fit to the cluster mass function within a radial bin $M_{\rm cl,c}$ and the maximum cluster mass within a radial bin $M_{\rm cl,max}$. The first mass-scales are averaged both azimuthally and over the bin, whereas the second ones are not, indicating they are equivalent to the individual cluster masses we use for the Milky Way. Looking into Fig.~\ref{fig:fig3-m83}, there is a slight disagreement between the predictions of our model and the maximum mass-scales in the first and second bin. This region is where the bar should have an effect, but given that the disagreement is within the errorbars we can conclude that cluster formation is only slightly affected by the bar.

\subsection{ZC406690} \label{sec:predictions-zc406690}

As a final test of our model, we use the high-redshift galaxy zC406690 ($z = 2.196$) from the zCOSMOS-SINFONI sample (see \citealt{forsterschreiber09,mancini11}). This choice is motivated by its average properties for a high-redshift galaxy (see e.g. \citealt{mancini11}, \citealt{tacconi13}), except for its high SFR, which is typical of a highly actively star-forming environment. Next to zC406690, there are two additional galaxies in the samples from \citet{genzel11} and \citet{tacconi13} with known clump masses and macroscopic observational parameters. However, zC406690 is the only galaxy that is classified as rotationally-dominated, which rules out the other galaxies as suitable targets for the application of our model.

\subsubsection{Prediction of the maximum mass-scales for clumps and stellar clusters} \label{subsec:pred-zc406690}

We now compare the predictions of our model to the observed clump masses from \citet{genzel11} derived using $H\alpha$ data obtained as part of the SINS GTO survey (\citealt{forsterschreiber09}) and the SINS/zCOSMOS ESO Large Program (see \citealt{mancini11}) of high-redshift galaxy kinematics carried out with SINFONI at the Very Large Telescope (VLT).

In order to properly describe the host galaxy gas disc, we use the properties derived from CO data obtained in the context of the PHIBSS survey with the Plateau de Bure millimeter Interferometer (PdBI, nowadays called NOEMA) and listed in Table 2 of \citet{tacconi13}. In particular, we use the circular velocity $V = 224\,\kms$, the optical half-mass radius $R_{1/2} = 6.3\,\kpc$, and the mean molecular gas surface density contained in the half-mass radius $\Sigma_{\rm g} (R\leq R_{1/2}) = 10^{2.52}\,\msun\,\pc^{-2}$.\footnote{The mean molecular gas surface density is determined as $\Sigma_{\rm g} (R\leq R_{1/2}) = M_{\rm mol-gas}/(\pi R_{1/2}^2)$, where the molecular mass of the cloud is derived from CO data using the standard Galactic CO-to-H$_2$ conversion factor of $X_{\rm CO} = 2\times 10^{20}\,\rm cm^{-2}\,(K\,\kms)^{-1}$ (\citealt{bolatto13}).} These global properties are then converted to radial profiles. In principle, \citet{genzel11} use the H$\alpha$ emission in conjunction with the gas depletion time to estimate a molecular gas surface density profile. However, while this method is intrinsically indirect, it also provides the {\it present} gas surface density profile, in which the clumps have already condensed. We are interested in the initial gas surface density profile, {\it before} the clumps have formed. We therefore assume that the gas initially followed an exponential profile with the same half-light radius as currently seen in the optical. The ratio between the half-mass radius and the scale radius in a two-dimensional exponential surface density profile is constant at $R_{1/2}/R_{\rm d}=1.678$, which allows us to express this profile as:
\be
\Sigma_{\rm g} (R) = \Sigma_0 \exp\left(-R/R_{\rm d}\right)
\ee
with
\be
\Sigma_0 = \dfrac{\Sigma_{\rm g}(R\leq R_{1/2})}{2}\dfrac{\left(R_{1/2}/R_{\rm d}\right)^2}{\left[1-\exp\left(-\frac{R_{1/2}}{R_{\rm d}}\right)\left(1+\frac{R_{1/2}}{R_{\rm d}}\right)\right]},
\ee
We assume that the rotation curve is flat at the quoted circular velocity and use a velocity dispersion of $\sigma=50~\kms$ to determine $Q$.

\begin{figure}
	\includegraphics[width=\hsize]{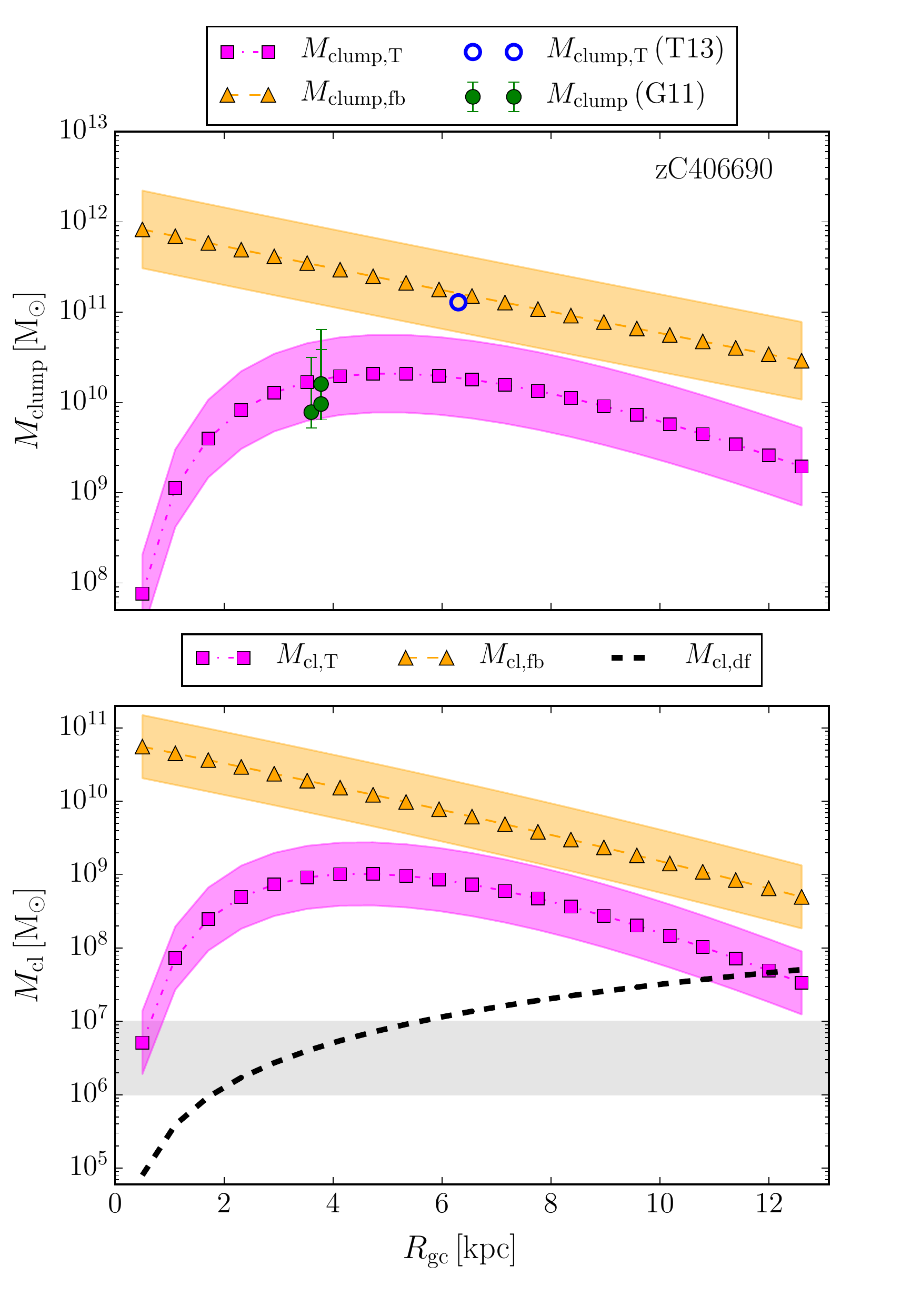}
	\caption{
		\label{fig:fig3-zc406690}
		Result of applying our shear-feedback hybrid model to zC406690 using CO data from the PHIBSS survey: (\textit{top}) maximum clump mass, and (\textit{bottom}) maximum cluster mass as a function of the galactocentric radius. The magenta squares correspond to the shear-limited mass-scales and the orange triangles correspond to the feedback-limited mass-scales. In the bottom panel, the dashed line indicates the maximum cluster mass that may survive dynamical friction over a typical major merger time-scale, i.e.~before the clusters would be redistributed by mergers. The observed clump masses from \citet{genzel11} are represented by the green filled dots with error bars. For comparison, the blue open circle shows the predicted shear-limited mass when using the global-average properties from Table 2 in \citet{tacconi13} rather than the radial profiles used here. The colour-shaded areas indicate a fiducial uncertainty range, assuming that the rotation curve and the gas surface density profile are known to an accuracy of $\sim0.04$~dex and $\sim0.13$~dex, respectively. The grey-shaded area in the bottom panel indicates the typical maximum mass-scale of globular clusters (see the text).}
\end{figure}

We thus apply the model described in Section~\ref{sec:model} to the obtained values for a radial range out to twice the optical half-mass radius. We determine the shear and feedback-limited maximum cloud and cluster mass-scales and we show them in Fig.~\ref{fig:fig3-zc406690}. Assuming typical uncertainties of $10$ per cent in the epicyclic frequency and $30$ per cent in the gas surface density, the uncertainty associated with our predictions is $\sim0.4\,\rm dex$.

We overplot the observed clump masses from \citet{genzel11} in the top panel with an error bar of a factor 3 (as indicated in their paper). We also plot the shear-limited Toomre mass determined directly from the galaxy-average properties listed in \citet{tacconi13}. Its large discrepancy relative to the predicted mass-scales using the exponential fit is due to the use of the mean surface density contained within a certain radius rather than the density at that radius, which is much lower. The predicted clump mass-scales by our model correspond to the shear-limited Toomre masses, and they are of the order of the observed clump masses. \citet{dekel13} demonstrate that stellar feedback is expected to produce steady outflows from the giant clumps in high-redshift galaxies, but they are not expected to be disrupted in the process. The authors argue that the clumps are expected to migrate to the centre to build-up the bulge in a short time-scale, although recent work suggests this process may be inhibited by short clump lifetimes (e.g.~\citealt{oklopcic17}). 

A consideration to be made is whether the larger mass-scales observed for the high-redshift clumps than for the local-Universe GMCs are result of the observing resolution, as recent studies suggest (e.g. \citealt{behrendt16}). The idea that high-redshift clumps may be analogous to the GMCs in the local Universe has already been suggested in previous studies (e.g.~\citealt{oklopcic17}). The massive clumps may fragment into smaller ones, which at low spatial resolution could appear as a single clump, but could also end up merging into a single clump. Because this paper focuses on the maximum mass-scales, we consider only the optimal case where, even if the clumps fragment, they will merge again into a single cloud.

The predicted cluster masses, shown in the bottom panel in Fig.~\ref{fig:fig3-zc406690}, are much larger than any cluster mass observed at $z=0$. We investigate the possibility of dynamical friction being the mechanism that destroys the more massive clusters. In the context of the \citet{kruijssen15b} globular cluster formation model, stellar clusters are disrupted by tidal shocks until they are redistributed to the halo during galaxy mergers or in accretion events (also see \citealt{kravtsov05}, \citealt{prieto08}, \citealt{rieder13}). Equation~(18) of \citet{kruijssen15b} shows the minimum cluster mass that will spiral in a time-scale shorter than a galaxy merging time, and thus, it will get destroyed by dynamical friction before it can escape to the halo,
\be
M_{\rm cl,df} = 10^6\,\msun \left(\dfrac{t_{\rm merge}}{2\,\gyr}\right)^{-1}\left(\dfrac{R}{2\,\kpc}\right)^2\left(\dfrac{V}{200\,\kms}\right),
\ee
\noindent
where $t_{\rm merge}$ is the galaxy merger time-scale, $R$ is the galactocentric radius of the stellar cluster and $V$ is the circular velocity. Given the galaxy’s stellar mass of $M_*\simeq 4\times10^{10}~\msun$, we infer a halo mass at $z\simeq2.2$ of $M_{\rm h}\simeq 2\times10^{12}~\msun$ through abundance matching \citep{moster13}. For such a halo mass, we use the relations from \citet{genel09} to estimate that the time-scale to merge with another halo of at least 1/3 its own mass\footnote{We follow the assumption of \citet{kruijssen15b} that this is a reasonable minimum mass ratio for driving the redistribution of the clusters.} is $t_{\rm merge}\simeq 3.5~\gyr$. We use the same flat rotation curve of $V=224\,\kms$ and obtain an expression that only depends on the galactocentric radius of the cluster:
\be
M_{\rm cl,df} = 3.2\times10^5\,\msun\left(\dfrac{R}{\kpc}\right)^2.
\ee
\noindent
We plot this relation in the bottom panel in Fig.~\ref{fig:fig3-zc406690}. Only the most massive clusters on the edge of the radial range considered ($R>12~\kpc$) may spiral in slowly enough to be ejected before they become part of the nucleus. Therefore, the minimum cluster mass for destruction by dynamical friction becomes the effective predictor for the maximum mass-scales  of globular clusters at $z=0$. Indeed, the predicted mass range matches the observed truncation mass of the globular cluster mass function (GCMF) of about $10^6$--$10^7~\msun$ \citep[e.g.][]{fall01,jordan07,kruijssen09b}. This result is in contradiction to the finding of \citet{jordan07} that dynamical friction does not drive the truncation of the GCMF. However, we note that the present work considers the observed conditions in real high-redshift galaxies, in which globular clusters must have formed, whereas \citet{jordan07} arrived at their conclusions by using the properties of the present-day galaxy population.

\subsubsection{Comparison to previous work on the maximum mass-scales of high-redshift clumps}

This is not the first time that the shear-limited Toomre mass is calculated for high-redshift galaxies, with the goal of determining the maximum clump mass. \citet{dekel09} derive a simplified expression for the Toomre mass that is based on global galaxy properties, by assuming that the angular velocity traces the gravitational potential (i.e.~the disc is rotation-dominated). This allows them to relate the Toomre clump mass to a fraction of the disc mass. We use this formalism to evaluate whether the sample from \citet{tacconi13} can be used to test our model.

\citet{dekel09} present a simple theoretical framework for massive galaxies at high redshift in which discs are considered to be self-gravitating and rotating objects subject to gravitational instabilities. Despite recent work calling into question the validity of using the linear Toomre analysis in the context of highly non-linear galaxies (\citealt{behrendt15}, \citealt{tamburello15}, \citealt{inoue16}), these concerns mainly relate to the subsequent fragmentation of the clumps. Their maximum mass-scales are still largely set by the balance between self-gravity, the centrifugal force and the turbulent pressure. Thus, the larger clumps are limited by shear and correspond to the Toomre mass, to which the results of the present work adds further support. Following a similar line of reasoning, \citet{dekel09} use the characteristic Toomre length to define the properties of the clumps. They use a number of assumptions to relate those properties with the radius and the mass of the disc component of the galaxy. We now re-evaluate these assumptions and use the high-redshift galaxies listed in \citet{tacconi13} to see if the parameter space studied is affected.

\citet{dekel09} define two quantities, $\delta$ and $\beta$, which correspond to the mass fraction of the disc and to the baryon mass fraction, respectively,
\be
\delta\equiv\dfrac{M_{\rm d}}{M_{\rm tot}} \leq \beta \equiv \dfrac{M_{\rm bar}}{M_{\rm tot}}
\label{eq:inequality-delta-beta}
\ee
\noindent
where $M_{\rm d}$ is the mass in the disc component within the disc radius $R_{\rm d}$, $M_{\rm bar}$ is the mass of the baryons in the disc and bulge within $R_{\rm d}$ and $M_{\rm tot}$ is the total mass taking into account the dark matter, the stellar and the disc masses within $R_{\rm d}$.

The inequality of equation~(\ref{eq:inequality-delta-beta}) implies that the disc mass has to be smaller than or equal to the baryonic mass, $M_{\rm d} \leq M_{\rm bar} \equiv \beta M_{\rm tot}$. \citet{dekel09} determine the total mass from the expression of the circular velocity at the disc radius $R_{\rm d}$,
\be
\left(\Omega R_{\rm d}\right)^2 = V^2 = \dfrac{G M_{\rm tot}}{R_{\rm d}} \rightarrow M_{\rm tot} = \dfrac{\Omega^2 R_{\rm d}^3}{G}
\label{eq:gravitational-mass}
\ee
\noindent
and assume a constant surface density within $R_{\rm d}$ to determine the disc mass, $M_{\rm d} = \pi\Sigma_{\rm tot}R_{\rm d}^2=\pi\Sigma_{\rm g} R_{\rm d}^2/f_{\rm g}$, where $f_{\rm g}=M_{\rm g}/(M_{\rm g}+M_*)$ is the gas fraction. Assuming a bulgeless disc (i.e.~$\delta=\beta$ and $M_{\rm d}=M_{\rm bar}$), this then allows \citet{dekel09} to write:
\be
\label{eq:mtoomre_simple}
M_{\rm T}\simeq \frac{\pi^2}{36}\beta^2 M_{\rm d} ,
\ee
with the express requirement that $0\leq\beta\leq1$.

By introducing the above expressions in the mass in equation~(\ref{eq:inequality-delta-beta}) and again assuming $\delta=\beta$, we can define the combination of the baryon and the gas fraction with its physical range:
\be
\dfrac{\pi\Sigma_{\rm g} R_{\rm d}^2}{f_{\rm g}} = \beta \dfrac{\Omega^2 R_{\rm d}^3}{G} \rightarrow 1 \geq  \beta f_{\rm g} = \dfrac{\pi G \Sigma_{\rm g}}{\Omega^2 R_{\rm d}} \geq 0
\label{eq:condition-beta}
\ee
\noindent

To test if the simplified expression for the maximum mass-scale holds, i.e.~if the product of the baryon fraction and gas fraction apply, we use the high-redshift galaxies listed in \citet{tacconi13}. We separate the galaxies into two subsamples depending on the kinematic classification done by \citet{tacconi13}; \textit{rotation} are galaxies with disc-like morphology and a velocity gradient in the CO data and \textit{dispersion} are disc galaxies without a velocity gradient (\textit{disk(A)} and \textit{disk(B)} in their classification). We use the circular velocity $V$, the half-light optical radius $R_{1/2}$, the gas fraction $f_{\rm g}$, and the mean gas surface density within the half-light optical radius $\Sigma_{\rm g} (r<R_{1/2})$ listed in their Table 2. We determine the angular velocity at the half-light radius as $\Omega_{\rm obs} = V/R_{1/2}$. We restrict the galaxies to be in the parameter space studied in Fig.~\ref{fig:fig1-general}.

\begin{figure}
	\centering
	\includegraphics[width=\hsize]{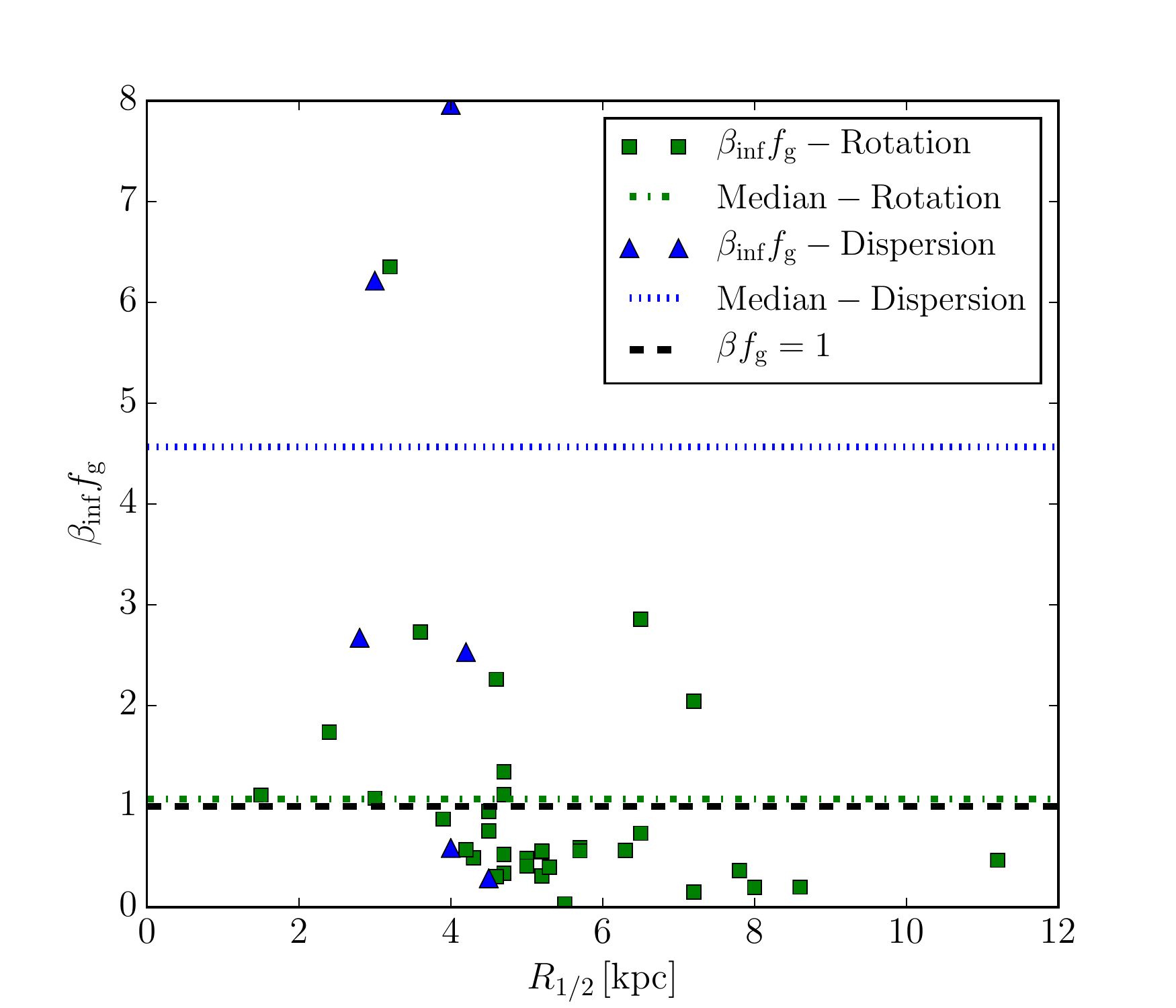}
	\caption{
		\label{fig:beta-radius}
		Product of the baryon fraction $\beta$ and the gas fraction $f_{\rm g}$ that would be implied by using the simplified expression of the Toomre mass from \citet{dekel09} as a function of the half-light optical radius $R_{1/2}$ for two sub-samples of high-redshift galaxies from \citet{tacconi13}, i.e.~dispersion or rotation-dominated. We overplot the median for each subsample (green line for rotation, blue line for dispersion-dominated) and the unity line (black dashed line), which represents the maximum allowed value of $\beta f_{\rm g}$.}
\end{figure}

With these values ($\Omega_{\rm obs}$, $\Sigma_{\rm g}$, $f_{\rm g}$, and $R_{1/2}$) we determine the value of the product of the baryon and gas fraction $\beta_{\rm inf} f_{\rm g}$ required by the combination of the angular velocity and the baryonic surface density for each subsample of galaxies using equation~(\ref{eq:condition-beta}). We show it as a function of the half-light radius of the galaxy in Fig.~\ref{fig:beta-radius}. We overplot the median of each subsample, as well as the maximum value for $\beta_{\rm inf} f_{\rm gas}= 1$. The galaxies dominated by dispersion show a larger dispersion in $\beta f_{\rm gas}$, whereas the rotation-dominated galaxies have a significantly lower ratio between the gas mass and the dynamical mass, to such an extent that the median is consistent with the allowed range of $\beta f_{\rm g}\leq 1$. This comparison shows that dispersion-dominated galaxies have measured values of $\beta f_{\rm g}$ that are well in excess of the allowed range, which means that the \citet{dekel09} expression for the Toomre mass from equation~(\ref{eq:mtoomre_simple}) substantially overestimates the mass if the observed inferred $\beta f_{\rm g}$ are used.

We emphasise that the bias induced by unphysical values of $\beta f_{\rm g}$ is not exclusive to the simplified \citet{dekel09} expression for the Toomre mass, but is more immediately obvious in their formulation. Fundamentally, the bias arises because the inferred angular velocity in dispersion-dominated systems underestimates the depth of the gravitational potential and, hence, the total galaxy mass. This underestimate of $\Omega$ affects equation~(\ref{eq:mtoomre_simple}) in the same way as it affects the formal expression in equation~(\ref{eq:toomremass}) -- both effectively scale with $\Omega^{-4}$.

\begin{figure*}
	\centering
	\includegraphics[width=\hsize]{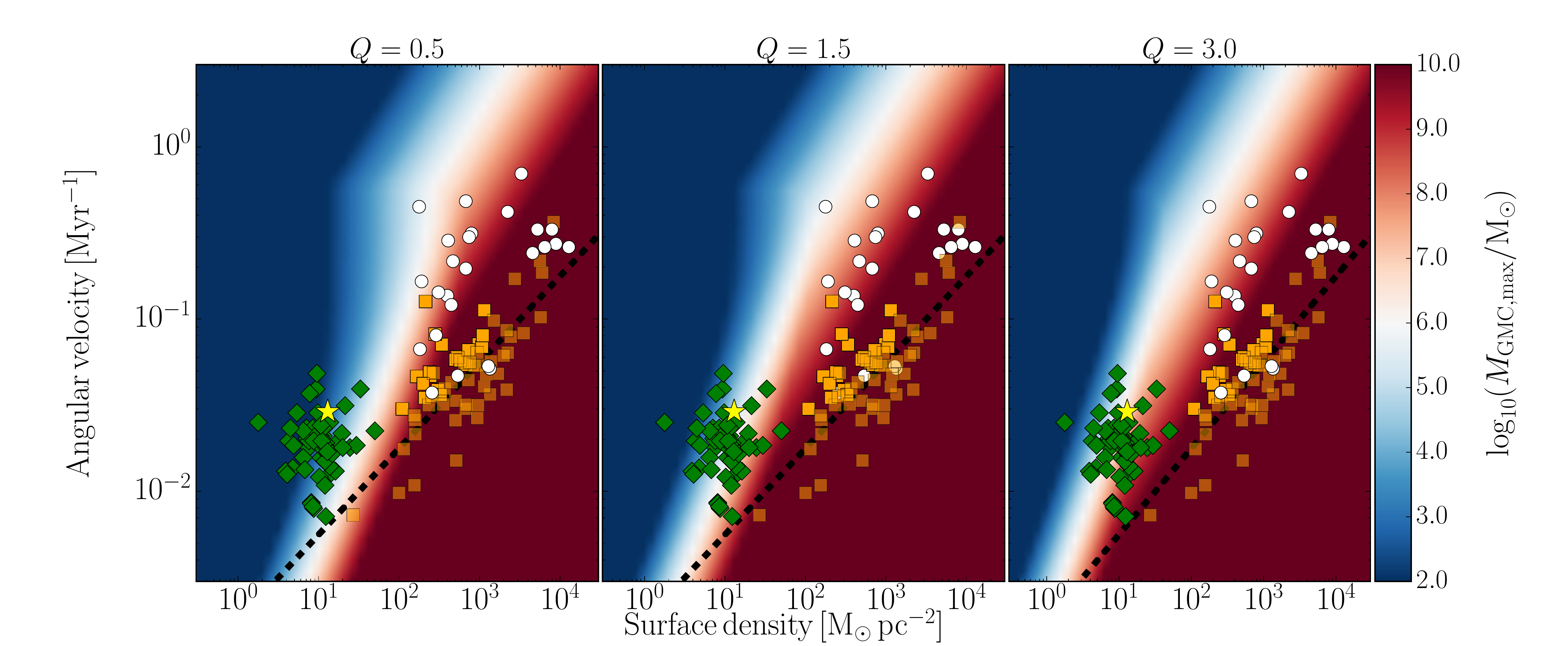}
	\caption{
		\label{fig:parameter-space}
		Maximum cloud mass as a function of the angular velocity and the surface density for three values of the stability parameter $Q$. We overplot observational data from four galactic environments as in Fig.~\ref{fig:fig1-general}. We shade the high-redshift galaxies that require baryon fractions larger than unity when using the simplified expression of the Toomre mass from \citet{dekel09}. The black dashed line indicates a required baryon fraction of unity for the mean disc size in the galaxy sample of \citet{tacconi13}.}
\end{figure*}

With this result we can revisit the condition on $\beta f_{\rm g}\leq1$ and obtain a condition for the angular velocity that isolates the part of parameter space in which the observed galaxy properties can be used reliably as input for our model:
\be
\dfrac{\pi G \Sigma_{\rm g}}{\Omega^2 R_{\rm d}} \leq 1 \rightarrow \Omega \geq \sqrt{\dfrac{\pi G \Sigma_{\rm g}}{R_{\rm d}}},
\label{eq:conditionOmega}
\ee
\noindent
where we use the mean galaxy size of the high-redshift sample for $R_{\rm d}$. This is a representative radius, because Fig.~\ref{fig:beta-radius} shows that the product $\beta f_{\rm g}$ has no strong radius dependence. In order to visualize better this condition, we plot it in a modification of Fig.~\ref{fig:fig1-ratioMgmc} that is shown in Fig.~\ref{fig:parameter-space}. We include the same galactic environments discussed in Section~\ref{sec:results}, but we shade the high-redshift galaxies for which the inferred low angular velocities imply a product of the baryon fraction and the gas fraction in excess of unity. These galaxies have gas masses exceeding the total mass needed to maintain their angular velocity.

The condition for the angular velocity clearly separates the high-redshift galaxy sample into two sub-samples; above the line, they fulfill the condition of having a gas-to-total mass fraction $\beta f_{\rm g}$ smaller than unity, but below the line this condition is violated. We find 44 high-redshift galaxies out of the total sample of 65 in \citet{tacconi13} that fall into this regime. The large implied gas-to-total mass fractions indicate that the observed angular velocity ($\Omega_{\rm obs}$) is smaller than the angular velocity determined from the surface density (equation~(\ref{eq:gravitational-mass}) and assuming $M_{\rm tot} = \pi R^2 \Sigma_{\rm g} f_{\rm g}$). In view of these results, we caution that only those high-redshift galaxies with implied gas-to-total mass fractions smaller than unity should be used to test the predictions of our model.

\section{Conclusions} \label{sec:concl}

We present a simple, self-consistent analytical model to determine the maximum mass of GMCs and stellar clusters as a function of the galactic environment. The model develops the idea that these maximum masses can be limited by shear and stellar feedback. In environments with low shear and low gas surface densities, feedback is expected to proceed more rapidly than the free-fall times of shear-limited GMCs, thus giving time to the massive stars to disrupt the cloud before the collapse has finished. Taking this into account, our model predicts smaller masses for both GMCs and stellar clusters in those environments.

We have explored the parameter space formed by $Q$, $\Sigma_{\rm g}$ and $\Omega$ and we predict that GMC and cluster masses are feedback-limited if both the galactic gas surface density $\Sigma_{\rm g} \leq 100\,\msun\pc^{-2}$ and the angular velocity $\Omega \leq 0.6\,\myr^{-1}$, assuming a typical $Q= 1.5$. For larger values of gas surface density and angular velocity, the masses become shear-limited. For lower (higher) values of $Q$, the limit for the gas surface density shifts towards higher (lower) values ($\Sigma_{\rm g} \leq 650\,\msun\pc^{-2}$ for $Q = 0.5$ and $\Sigma_{\rm g} \leq 45\,\msun\pc^{-2}$ for $Q = 3$), whereas the limit for the angular velocity does not change.

We also find that the region affected by feedback in our model encompasses the solar neighbourhood and the discs of local-Universe spiral galaxies described in \citet{kennicutt98}. On the other hand, the high-redshift galaxies from \citet{tacconi13} and the circumnuclear starbusts of \citet{kennicutt98} reside in the shear-limited regime. This transition between a feedback-dominated and a shear-dominated regime explains why the clumps observed in those environments have higher masses than those in local galaxies (\citealt{elmegreen05}), even beyond the environmental dependence already expected if the mass-scales would only be set by the shear-limited Toomre mass.

There are a couple of limitations that should be kept in mind when applying our model. We assume a differentially rotating disc in hydrostatic equilibrium. External compression of the material caused by the presence of substructure in the disc (such as bars, arms, or galaxy mergers) may yield larger mass-scales than those predicted by the model. At the same time, the model bases its predictions on the current gas conditions, so the cluster sample to which it has to be compared has to correspond to an age range where the gas surface density has remained stable. Last, the predictions of the model for clusters forming in high-redshift environments carry the caveat that such massive objects may not survive until the present day due to dynamical friction towards the bulge of the galaxy.

We compare our predictions as a function of the galactocentric radius for the Milky Way, the disc galaxy M31, the grand-design spiral galaxy M83, and the high-redshift galaxy zC406690 with the observed GMC, clump, and cluster masses in these galaxies. As in the numerical simulations by \citet{oklopcic17} (see also \citealt{soto17} for an observational study), we find that the molecular clumps become less massive with decreasing redshift; in the context of our analytical model, this happens because the galaxies become more extended and less gas-rich, driving them into the feedback-limited regime.

For the Milky Way, the key new ingredient of our model, i.e.~that the cloud can be disrupted by stellar feedback due to the feedback time-scale being smaller than the free-fall time, is applicable for galactocentric radii of $R \geq 4.3$~kpc. Observations indicate a roughly constant mass of $M_{\rm cl, max}\sim10^4\,\msun$ across all galactocentric radii, as our model predicts. For the CMZ and the inner part of the galaxy, the GMC masses are equal to the shear-limited Toomre mass. Our hybrid mass predictions correctly reproduce the maximum observed cloud mass in the CMZ and the solar neighbourhood, as well as the observed cluster masses at all radii, except for those located at the end of the bar. The compression of the gas at the end of the bar may aid the accumulation of a larger mass reservoir than would normally have been possible against shear and feedback. 

In M31, the feedback-timescale is smaller than the two-dimensional free-fall time-scale for galactocentric radii $R\geq8.4\,\kpc$. Our predictions for the maximum GMC mass agrees with the observed maximum cloud mass reported in \citet{schruba17}. However, our predictions based on the present day gas surface density do not reproduce the truncation mass reported in \citet{johnson17}. We use the spatially-resolved star formation history of M31 to show that this discrepancy is plausibly explained by the fact that the most massive clusters are older than $100\,\myr$ and formed under higher gas densities than currently observed. As opposed to the Milky Way and M31, the maximum mass-scales in M83 fall in the shear-limited regime of our model. Again, the predicted cloud and cluster mass-scales agree well with the observed masses at all galactocentric radii.

\citet{johnson17} have recently proposed a relation between the high-mass end of the young cluster mass function and the galaxy-averaged star formation rate surface density. In the context of our model, their $M_{\rm c}$--$\Sigma_{\rm SFR}$ relation does not account for the steep dependence of the Toomre mass on the epicyclic frequency ($M_{\rm T}\propto\kappa^{-4}$). We therefore interpret the relation proposed by \citet{johnson17} as the result of selecting environments at similar epicyclic frequencies, which hides the $\kappa$ dependence and allows the residual effect to be absorbed into $\Sigma_{\rm SFR}$. This happens because $\Sigma_{\rm SFR}$ is covariant with $\kappa$ due to the radial decrease of both quantities in nearly all galaxies. In principle, the dependence on $\kappa$ should affect the maximum mass-scale everywhere, but most noticeably so in shear-dominated environments like the CMZ or M83. We use the CMZ as a test case to compare the prediction from the $M_{\rm c}$--$\Sigma_{\rm SFR}$ relation to that of our model, as well as to the observed maximum cluster masses. We consider the CMZ within $|l|\leq1^\circ$, corresponding to $R\leq140\,\pc$, where the SFR is ${\rm SFR} = 0.1\,\msun\yr^{-1}$ \citep{longmore13,barnes17}. With these values, we obtain a SFR surface density of $\Sigma_{\rm SFR}=1.6\,\msun\,\yr^{-1}\,\kpc^{-2}$, for which the relation between $M_{\rm c}$ and $\Sigma_{\rm SFR}$ suggested by \citet{johnson17} yields a Schechter characteristic mass of $M_{\rm c} \simeq 1\times10^7\,\msun$. By contrast, our model predicts a shear-limited maximum cluster mass of $M_{\rm cl,max}\simeq3\times10^4\,\msun$. These predictions differ by nearly three orders of magnitude. If we now compare them to the most massive clusters in the CMZ, which are the Arches and Quintuplet clusters with masses of $M\simeq2\times10^{4}\,\msun$ \citep{portegieszwart10}, we find that our model agrees very well with the observed cluster mass-scales, whereas the $M_{\rm c}$--$\Sigma_{\rm SFR}$ relation strongly overpredicts the cluster masses. This is also reflected by the maximum cloud masses observed in the CMZ -- the GMCs on the 100-pc stream have masses of $\sim10^5~\msun$ \citep{walker15}, well below the maximum {\it cluster} mass one would infer from the proposed $M_{\rm c}$--$\Sigma_{\rm SFR}$ relation.\footnote{One could argue that the total SFR in the CMZ is so low that the maximum cluster mass is set by the size-of-sample effect, which would mean that the physical maximum cluster mass is larger than observed. Fortunately, this effect can be quantified using Fig.~1 of \citet{gieles09}, who show the mass of the most massive cluster given the true maximum cluster mass as a function of the total mass of the cluster population $\Gamma\cdot{\rm SFR}\cdot\Delta t$, where $\Delta t$ is the age range of the clusters. For the observed ${\rm SFR} = 0.1\,\msun\yr^{-1}$, the expected CFE of $\Gamma\sim50$~per~cent \citep{kruijssen14b}, and a putative cluster age range of $\Delta t=10~\myr$, we find a total cluster population mass of $\Gamma\cdot{\rm SFR}\cdot\Delta t=5\times10^5~\msun$. For a Schechter-type initial cluster mass function with commonly-adopted slope $\beta=-2$, we see that the observed masses of the most massive clusters in the CMZ imply a maximum mass-scale of a few $10^4~\msun$, and certainly $<10^5~\msun$. This result is consistent with our prediction. Note that we are ignoring the suggestion that the SFR in the CMZ was even higher when the Arches and Quintuplet clusters formed \citep{krumholz17}, which would imply an underlying maximum mass even closer to the observed cluster masses. The size-of-sample effect is therefore not dominant in setting the maximum cluster masses observed in the CMZ, as expected from the fact that the clouds have masses of only $\sim10^5~\msun$ -- the material for forming more massive clusters is not available within the tidal limit.} This example highlights the importance of accounting for differential rotation and shear when predicting the maximum mass-scales for molecular clouds and stellar clusters.

The fact that the observed cluster masses agree with the feedback-limited cluster mass-scales predicted by our model for a significant subset of local-Universe environments suggests that the most massive GMCs are gravitationally collapsing (e.g. fig. 17 in \citealt{mivilledeschenes17}). Otherwise, the clusters forming within that cloud would not be able to continue their assembly into one object after the feedback has blown out the gas. Therefore, an important implication of our results is that the formation time-scale of both GMCs and high-redshift clumps has an upper limit set by the two-dimensional free-fall time at the Toomre length (also see \citealt{jeffreson17}). By adding feedback to the normal Toomre analysis, we obtain a formation time set by the minimum of the two-dimensional free-fall time in equation~(\ref{eq:tff-2D}) and the feedback time-scale in equation~(\ref{eq:tfbgmc}). We note that support against collapse by shear can significantly increase these time-scales, but only if $Q\ga3$ \citep{jeffreson17}.

The clump and cluster mass-scales in the high-redshift galaxy zC406690 are also limited by shear. The predictions of our model for the maximum clump masses agree with the observed clump masses from \citet{genzel11}. Our model predicts very large maximum cluster masses, but we demonstrate that, across the radial range covered, dynamical friction would cause these clusters to spiral into the nucleus before they might be scattered out into the halo by a major merger. Dynamical friction thus limits the maximum mass-scales of long-lived globular clusters at $z=0$. We predict globular clusters formed in zC406690 should not exceed $10^6$--$10^7~\msun$, which agrees with the observed truncation mass of the GCMF \citep{jordan07,kruijssen09b}. 

We use a commonly-used, simplified expression based on macroscopic galaxy properties \citep{dekel09} and find that the properties of observed, dispersion-dominated galaxies \citep{tacconi13} have strongly underestimated angular velocities, which leads to gas-to-total mass fractions in excess of unity and correspondingly overestimated maximum clump masses. We use this example to illustrate that the observed properties of galaxies are sometimes inconsistent. Only those high-redshift galaxies with implied gas-to-total mass fractions smaller than unity should be used to test the predictions of our model.

The shear-limited clump mass-scales predicted by our hybrid model for zC406690 are also confirmed by the results in \citet{fisher17b}, which appeared while the present paper was in review. They use the DYNAMO sample of low-redshift ($z\sim0.1$) analogues to clumpy high-redshift galaxies \citep{green14} to show that the clump sizes correspond roughly to the most shear-unstable scale, which corresponds to half the Toomre length. This supports the hypothesis that massive star-forming clumps in turbulent discs are the result of gravitational instabilities. In the context of our model, this result is to be expected, because the DYNAMO galaxies considered by \citet{fisher17} have typical surface densities and angular velocities of $\Sigma\sim400~\msun~\pc^{-2}$ and $\Omega\sim0.09~\myr^{-1}$, respectively,\footnote{The surface density is obtained by writing $\Sigma=f_{\rm gas}M_{\rm star}/2(1-f_{\rm gas})\pi R_{1/2}^2$, where the stellar mass ($M_{\rm star}\sim3\times10^{10}~\msun$), gas fraction ($f_{\rm gas}\sim0.3$), and half-light radius ($R_{1/2}\sim2.3~\kpc$) were determined for the first seven galaxies in Table~1 of \citet{fisher17b}. The angular velocity is obtained by writing $\Omega=V/R_{1/2}$, where $V\sim210~\kms$ is taken to be the average for the same seven galaxies in Table~1 of \citet{fisher17}.} which places them firmly in the shear-limited part of parameter space in Figure~\ref{fig:fig1-ratioT} for $Q>0.5$ and on the limit between both regimes for $Q\sim0.5$.

In the context of galaxy formation, our model predicts an evolution of the maximum GMC (or clump) and stellar cluster (or globular cluster) mass with galactic environment and thus with redshift. As galaxies grow, they become less gas-rich (\citealt{tacconi10}) and less compact due to their inside-out growth (\citealt{delucia08}, \citealt{allen17}), implying that, over time, galaxies evolve to the feedback-limited regime in the parameter space of our model. We demonstrate that this prediction agrees with the observed GMC and cluster masses for four different galactic environments.

It is particularly important to consider the evolution of the mass-scales with the environment in galaxy formation models that are aimed at reproducing the masses of molecular clouds and stellar clusters until $z = 0$. For instance, our results imply that the simple (star) particle-tagging techniques that are commonly used to assign globular clusters to the output of galaxy formation simulations result in biased samples of candidate globular clusters. Studies aiming to model the origin of the globular cluster population, i.e.~the most massive clusters in the Universe, must account for the variation with time and environment of the maximum cloud and cluster mass-scales.

\section*{Acknowledgements}
 We thank the anonymous referee for their helpful comments and suggestions. We thank Pamela Freeman and Erik Rosolowsky for providing the data from \citet{freeman17} in electronic form, Andreas Schruba for providing the gas surface profiles from \citet{schruba17} and Cliff Johnson for sharing the manuscript of \citet{johnson17} ahead of publication, as well as for helpful discussions. We acknowledge helpful discussions with Joel Pfeffer, as well as with the members of the MUSTANG group.
MRC would like to thank to the International Max Planck Research School for Astronomy and Cosmic Physics at the University of Heidelberg (IMPRS-HD) for the funding for this work. JMDK gratefully acknowledges support in the form of an Emmy Noether Research Group from the Deutsche Forschungsgemeinschaft (DFG), grant number KR4801/1-1.

\bibliographystyle{mnras}
\bibliography{mrcbib}

\begin{thebibliography}{}
\makeatletter
\relax
\def\mn@urlcharsother{\let\do\@makeother \do\$\do\&\do\#\do\^\do\_\do\%\do\~}
\def\mn@doi{\begingroup\mn@urlcharsother \@ifnextchar [ {\mn@doi@}
  {\mn@doi@[]}}
\def\mn@doi@[#1]#2{\def\@tempa{#1}\ifx\@tempa\@empty \href
  {http://dx.doi.org/#2} {doi:#2}\else \href {http://dx.doi.org/#2} {#1}\fi
  \endgroup}
\def\mn@eprint#1#2{\mn@eprint@#1:#2::\@nil}
\def\mn@eprint@arXiv#1{\href {http://arxiv.org/abs/#1} {{\tt arXiv:#1}}}
\def\mn@eprint@dblp#1{\href {http://dblp.uni-trier.de/rec/bibtex/#1.xml}
  {dblp:#1}}
\def\mn@eprint@#1:#2:#3:#4\@nil{\def\@tempa {#1}\def\@tempb {#2}\def\@tempc
  {#3}\ifx \@tempc \@empty \let \@tempc \@tempb \let \@tempb \@tempa \fi \ifx
  \@tempb \@empty \def\@tempb {arXiv}\fi \@ifundefined
  {mn@eprint@\@tempb}{\@tempb:\@tempc}{\expandafter \expandafter \csname
  mn@eprint@\@tempb\endcsname \expandafter{\@tempc}}}

\bibitem[\protect\citeauthoryear{{Adamo}, {Kruijssen}, {Bastian}, {Silva-Villa}
   \& {Ryon}}{{Adamo} et~al.}{2015}]{adamo15b}
{Adamo} A.,  {Kruijssen} J.~M.~D.,  {Bastian} N.,  {Silva-Villa} E.,   {Ryon}
  J.,  2015, \mn@doi [\mnras] {10.1093/mnras/stv1203}, \href
  {http://adsabs.harvard.edu/abs/2015MNRAS.452..246A} {452, 246}

\bibitem[\protect\citeauthoryear{{Allen} et~al.,}{{Allen}
  et~al.}{2017}]{allen17}
{Allen} R.~J.,  et~al., 2017, \mn@doi [\apjl] {10.3847/2041-8213/834/2/L11},
  \href {http://adsabs.harvard.edu/abs/2017ApJ...834L..11A} {834, L11}

\bibitem[\protect\citeauthoryear{{Barnes}, {Longmore}, {Battersby}, {Bally},
  {Kruijssen}, {Henshaw}  \& {Walker}}{{Barnes} et~al.}{2017}]{barnes17}
{Barnes} A.~L.,  {Longmore} S.~N.,  {Battersby} C.~D.,  {Bally} J.,
  {Kruijssen} J.~M.~D.,  {Henshaw} J.~D.,   {Walker} D.~L.,  2017,
  \mnras~submitted

\bibitem[\protect\citeauthoryear{{Bastian}}{{Bastian}}{2008}]{bastian08}
{Bastian} N.,  2008, \mn@doi [\mnras] {10.1111/j.1365-2966.2008.13775.x}, \href
  {http://adsabs.harvard.edu/abs/2008MNRAS.390..759B} {390, 759}

\bibitem[\protect\citeauthoryear{{Bastian} et~al.,}{{Bastian}
  et~al.}{2012}]{bastian12}
{Bastian} N.,  et~al., 2012, \mn@doi [\mnras]
  {10.1111/j.1365-2966.2011.19909.x}, \href
  {http://adsabs.harvard.edu/abs/2012MNRAS.419.2606B} {419, 2606}

\bibitem[\protect\citeauthoryear{{Behrendt}, {Burkert}  \&
  {Schartmann}}{{Behrendt} et~al.}{2015}]{behrendt15}
{Behrendt} M.,  {Burkert} A.,   {Schartmann} M.,  2015, \mn@doi [\mnras]
  {10.1093/mnras/stv027}, \href
  {http://adsabs.harvard.edu/abs/2015MNRAS.448.1007B} {448, 1007}

\bibitem[\protect\citeauthoryear{{Behrendt}, {Burkert}  \&
  {Schartmann}}{{Behrendt} et~al.}{2016}]{behrendt16}
{Behrendt} M.,  {Burkert} A.,   {Schartmann} M.,  2016, \mn@doi [\apjl]
  {10.3847/2041-8205/819/1/L2}, \href
  {http://adsabs.harvard.edu/abs/2016ApJ...819L...2B} {819, L2}

\bibitem[\protect\citeauthoryear{{Bigiel}, {Leroy}, {Walter}, {Brinks}, {de
  Blok}, {Madore}  \& {Thornley}}{{Bigiel} et~al.}{2008}]{bigiel08}
{Bigiel} F.,  {Leroy} A.,  {Walter} F.,  {Brinks} E.,  {de Blok} W.~J.~G.,
  {Madore} B.,   {Thornley} M.~D.,  2008, \mn@doi [\aj]
  {10.1088/0004-6256/136/6/2846}, \href
  {http://adsabs.harvard.edu/abs/2008AJ....136.2846B} {136, 2846}

\bibitem[\protect\citeauthoryear{{Bland-Hawthorn} \&
  {Gerhard}}{{Bland-Hawthorn} \& {Gerhard}}{2016}]{blandhawthorn16}
{Bland-Hawthorn} J.,  {Gerhard} O.,  2016, \mn@doi [\araa]
  {10.1146/annurev-astro-081915-023441}, \href
  {http://adsabs.harvard.edu/abs/2016ARA%26A..54..529B} {54, 529}

\bibitem[\protect\citeauthoryear{{Bolatto}, {Wolfire}  \& {Leroy}}{{Bolatto}
  et~al.}{2013}]{bolatto13}
{Bolatto} A.~D.,  {Wolfire} M.,   {Leroy} A.~K.,  2013, \mn@doi [\araa]
  {10.1146/annurev-astro-082812-140944}, \href
  {http://adsabs.harvard.edu/abs/2013ARA%26A..51..207B} {51, 207}

\bibitem[\protect\citeauthoryear{{Braun}, {Thilker}, {Walterbos}  \&
  {Corbelli}}{{Braun} et~al.}{2009}]{braun09}
{Braun} R.,  {Thilker} D.~A.,  {Walterbos} R.~A.~M.,   {Corbelli} E.,  2009,
  \mn@doi [\apj] {10.1088/0004-637X/695/2/937}, \href
  {http://adsabs.harvard.edu/abs/2009ApJ...695..937B} {695, 937}

\bibitem[\protect\citeauthoryear{{Burkert} \& {Hartmann}}{{Burkert} \&
  {Hartmann}}{2004}]{burkert04}
{Burkert} A.,  {Hartmann} L.,  2004, \mn@doi [\apjl] {10.1086/424895}, \href
  {http://adsabs.harvard.edu/abs/2004ApJ...616..288B} {616, 288}

\bibitem[\protect\citeauthoryear{{Caldwell}, {Harding}, {Morrison}, {Rose},
  {Schiavon}  \& {Kriessler}}{{Caldwell} et~al.}{2009}]{caldwell09}
{Caldwell} N.,  {Harding} P.,  {Morrison} H.,  {Rose} J.~A.,  {Schiavon} R.,
  {Kriessler} J.,  2009, \mn@doi [\aj] {10.1088/0004-6256/137/1/94}, \href
  {http://adsabs.harvard.edu/abs/2009AJ....137...94C} {137, 94}

\bibitem[\protect\citeauthoryear{{Colombo} et~al.,}{{Colombo}
  et~al.}{2014}]{colombo14}
{Colombo} D.,  et~al., 2014, \mn@doi [\apj] {10.1088/0004-637X/784/1/3}, \href
  {http://adsabs.harvard.edu/abs/2014ApJ...784....3C} {784, 3}

\bibitem[\protect\citeauthoryear{{Corbelli}, {Lorenzoni}, {Walterbos}, {Braun}
  \& {Thilker}}{{Corbelli} et~al.}{2010}]{corbelli10}
{Corbelli} E.,  {Lorenzoni} S.,  {Walterbos} R.,  {Braun} R.,   {Thilker} D.,
  2010, \mn@doi [\aap] {10.1051/0004-6361/200913297}, \href
  {http://adsabs.harvard.edu/abs/2010A%26A...511A..89C} {511, A89}

\bibitem[\protect\citeauthoryear{{De Lucia} \& {Helmi}}{{De Lucia} \&
  {Helmi}}{2008}]{delucia08}
{De Lucia} G.,  {Helmi} A.,  2008, \mn@doi [\mnras]
  {10.1111/j.1365-2966.2008.13862.x}, \href
  {http://adsabs.harvard.edu/abs/2008MNRAS.391...14D} {391, 14}

\bibitem[\protect\citeauthoryear{{Dekel} \& {Krumholz}}{{Dekel} \&
  {Krumholz}}{2013}]{dekel13}
{Dekel} A.,  {Krumholz} M.~R.,  2013, \mn@doi [\mnras] {10.1093/mnras/stt480},
  \href {http://adsabs.harvard.edu/abs/2013MNRAS.432..455D} {432, 455}

\bibitem[\protect\citeauthoryear{{Dekel}, {Sari}  \& {Ceverino}}{{Dekel}
  et~al.}{2009}]{dekel09}
{Dekel} A.,  {Sari} R.,   {Ceverino} D.,  2009, \mn@doi [\apj]
  {10.1088/0004-637X/703/1/785}, \href
  {http://adsabs.harvard.edu/abs/2009ApJ...703..785D} {703, 785}

\bibitem[\protect\citeauthoryear{{Elmegreen}}{{Elmegreen}}{2002}]{elmegreen02}
{Elmegreen} B.~G.,  2002, \mn@doi [\apj] {10.1086/342177}, \href
  {http://adsabs.harvard.edu/cgi-bin/nph-bib_query?bibcode=2002ApJ...577..206E&db_key=AST}
  {577, 206}

\bibitem[\protect\citeauthoryear{{Elmegreen}}{{Elmegreen}}{2010}]{elmegreen10}
{Elmegreen} B.~G.,  2010, \mn@doi [\apjl] {10.1088/2041-8205/712/2/L184}, \href
  {http://adsabs.harvard.edu/abs/2010ApJ...712L.184E} {712, L184}

\bibitem[\protect\citeauthoryear{{Elmegreen} \& {Elmegreen}}{{Elmegreen} \&
  {Elmegreen}}{2005}]{elmegreen05}
{Elmegreen} B.~G.,  {Elmegreen} D.~M.,  2005, \mn@doi [\apj] {10.1086/430514},
  \href {http://adsabs.harvard.edu/abs/2005ApJ...627..632E} {627, 632}

\bibitem[\protect\citeauthoryear{{Evans}}{{Evans}}{1999}]{evans99}
{Evans} II N.~J.,  1999, \mn@doi [\araa] {10.1146/annurev.astro.37.1.311},
  \href {http://adsabs.harvard.edu/abs/1999ARA%26A..37..311E} {37, 311}

\bibitem[\protect\citeauthoryear{{Fall} \& {Zhang}}{{Fall} \&
  {Zhang}}{2001}]{fall01}
{Fall} S.~M.,  {Zhang} Q.,  2001, \mn@doi [\apj] {10.1086/323358}, \href
  {http://adsabs.harvard.edu/cgi-bin/nph-bib_query?bibcode=2001ApJ...561..751F&db_key=AST}
  {561, 751}

\bibitem[\protect\citeauthoryear{{Fisher} et~al.,}{{Fisher}
  et~al.}{2017a}]{fisher17b}
{Fisher} D.~B.,  et~al., 2017a, \apjl~in~press, arXiv:1703.00458, \href
  {http://adsabs.harvard.edu/abs/2017arXiv170300458F} {}

\bibitem[\protect\citeauthoryear{{Fisher} et~al.,}{{Fisher}
  et~al.}{2017b}]{fisher17}
{Fisher} D.~B.,  et~al., 2017b, \mn@doi [\mnras] {10.1093/mnras/stw2281}, \href
  {http://adsabs.harvard.edu/abs/2017MNRAS.464..491F} {464, 491}

\bibitem[\protect\citeauthoryear{{F{\"o}rster Schreiber} et~al.,}{{F{\"o}rster
  Schreiber} et~al.}{2009}]{forsterschreiber09}
{F{\"o}rster Schreiber} N.~M.,  et~al., 2009, \mn@doi [\apj]
  {10.1088/0004-637X/706/2/1364}, \href
  {http://adsabs.harvard.edu/abs/2009ApJ...706.1364F} {706, 1364}

\bibitem[\protect\citeauthoryear{{Fouesneau} et~al.,}{{Fouesneau}
  et~al.}{2014}]{fouesneau14}
{Fouesneau} M.,  et~al., 2014, \mn@doi [\apj] {10.1088/0004-637X/786/2/117},
  \href {http://adsabs.harvard.edu/abs/2014ApJ...786..117F} {786, 117}

\bibitem[\protect\citeauthoryear{{Freeman}, {Rosolowsky}, {Kruijssen},
  {Bastian}  \& {Adamo}}{{Freeman} et~al.}{2017}]{freeman17}
{Freeman} P.,  {Rosolowsky} E.,  {Kruijssen} J.~M.~D.,  {Bastian} N.,   {Adamo}
  A.,  2017, \mnras~in~press, arXiv:1702.07728, \href
  {http://adsabs.harvard.edu/abs/2017arXiv170207728F} {}

\bibitem[\protect\citeauthoryear{{Frerking}, {Langer}  \& {Wilson}}{{Frerking}
  et~al.}{1982}]{frerking82}
{Frerking} M.~A.,  {Langer} W.~D.,   {Wilson} R.~W.,  1982, \mn@doi [\apj]
  {10.1086/160451}, \href {http://adsabs.harvard.edu/abs/1982ApJ...262..590F}
  {262, 590}

\bibitem[\protect\citeauthoryear{{Genel}, {Genzel}, {Bouch{\'e}}, {Naab}  \&
  {Sternberg}}{{Genel} et~al.}{2009}]{genel09}
{Genel} S.,  {Genzel} R.,  {Bouch{\'e}} N.,  {Naab} T.,   {Sternberg} A.,
  2009, \mn@doi [\apj] {10.1088/0004-637X/701/2/2002}, \href
  {http://adsabs.harvard.edu/abs/2009ApJ...701.2002G} {701, 2002}

\bibitem[\protect\citeauthoryear{{Genzel} et~al.,}{{Genzel}
  et~al.}{2011}]{genzel11}
{Genzel} R.,  et~al., 2011, \mn@doi [\apj] {10.1088/0004-637X/733/2/101}, \href
  {http://adsabs.harvard.edu/abs/2011ApJ...733..101G} {733, 101}

\bibitem[\protect\citeauthoryear{{Genzel} et~al.,}{{Genzel}
  et~al.}{2014}]{genzel14}
{Genzel} R.,  et~al., 2014, \mn@doi [\apj] {10.1088/0004-637X/785/1/75}, \href
  {http://adsabs.harvard.edu/abs/2014ApJ...785...75G} {785, 75}

\bibitem[\protect\citeauthoryear{{Gieles}}{{Gieles}}{2009}]{gieles09}
{Gieles} M.,  2009, \mn@doi [\mnras] {10.1111/j.1365-2966.2009.14473.x}, \href
  {http://adsabs.harvard.edu/abs/2009MNRAS.394.2113G} {394, 2113}

\bibitem[\protect\citeauthoryear{{Gieles}, {Larsen}, {Anders}, {Bastian}  \&
  {Stein}}{{Gieles} et~al.}{2006}]{gieles06a}
{Gieles} M.,  {Larsen} S.~S.,  {Anders} P.,  {Bastian} N.,   {Stein} I.~T.,
  2006, \mn@doi [\aap] {inserted by hand later}, 450, 129

\bibitem[\protect\citeauthoryear{{Green} et~al.,}{{Green}
  et~al.}{2014}]{green14}
{Green} A.~W.,  et~al., 2014, \mn@doi [\mnras] {10.1093/mnras/stt1882}, \href
  {http://adsabs.harvard.edu/abs/2014MNRAS.437.1070G} {437, 1070}

\bibitem[\protect\citeauthoryear{{Heiles} \& {Troland}}{{Heiles} \&
  {Troland}}{2003}]{heiles03}
{Heiles} C.,  {Troland} T.~H.,  2003, \mn@doi [\apj] {10.1086/367828}, \href
  {http://adsabs.harvard.edu/abs/2003ApJ...586.1067H} {586, 1067}

\bibitem[\protect\citeauthoryear{{Henshaw}, {Longmore}  \&
  {Kruijssen}}{{Henshaw} et~al.}{2016}]{henshaw16}
{Henshaw} J.~D.,  {Longmore} S.~N.,   {Kruijssen} J.~M.~D.,  2016, \mn@doi
  [\mnras] {10.1093/mnrasl/slw168}, \href
  {http://adsabs.harvard.edu/abs/2016MNRAS.463L.122H} {463, L122}

\bibitem[\protect\citeauthoryear{{Heyer}, {Krawczyk}, {Duval}  \&
  {Jackson}}{{Heyer} et~al.}{2009}]{heyer09}
{Heyer} M.,  {Krawczyk} C.,  {Duval} J.,   {Jackson} J.~M.,  2009, \mn@doi
  [\apj] {10.1088/0004-637X/699/2/1092}, \href
  {http://adsabs.harvard.edu/abs/2009ApJ...699.1092H} {699, 1092}

\bibitem[\protect\citeauthoryear{{Hughes} et~al.,}{{Hughes}
  et~al.}{2013}]{hughes13}
{Hughes} A.,  et~al., 2013, \mn@doi [\apj] {10.1088/0004-637X/779/1/46}, \href
  {http://adsabs.harvard.edu/abs/2013ApJ...779...46H} {779, 46}

\bibitem[\protect\citeauthoryear{{Inoue}, {Dekel}, {Mandelker}, {Ceverino},
  {Bournaud}  \& {Primack}}{{Inoue} et~al.}{2016}]{inoue16}
{Inoue} S.,  {Dekel} A.,  {Mandelker} N.,  {Ceverino} D.,  {Bournaud} F.,
  {Primack} J.,  2016, \mn@doi [\mnras] {10.1093/mnras/stv2793}, \href
  {http://adsabs.harvard.edu/abs/2016MNRAS.456.2052I} {456, 2052}

\bibitem[\protect\citeauthoryear{{Jeffreson} \& {Kruijssen}}{{Jeffreson} \&
  {Kruijssen}}{2017}]{jeffreson17}
{Jeffreson} S.~M.~R.,  {Kruijssen} J.~M.~D.,  2017, \mnras~submitted

\bibitem[\protect\citeauthoryear{{Johnson} et~al.,}{{Johnson}
  et~al.}{2017}]{johnson17}
{Johnson} L.~C.,  et~al., 2017, \apj~in~press, arXiv:1703.10312, \href
  {https://arxiv.org/abs/1703.10312} {}

\bibitem[\protect\citeauthoryear{{Jord{\'a}n} et~al.,}{{Jord{\'a}n}
  et~al.}{2007}]{jordan07}
{Jord{\'a}n} A.,  et~al., 2007, \mn@doi [\apjs] {10.1086/516840}, \href
  {http://adsabs.harvard.edu/abs/2007ApJS..171..101J} {171, 101}

\bibitem[\protect\citeauthoryear{{Kennicutt}}{{Kennicutt}}{1998}]{kennicutt98}
{Kennicutt} Jr. R.~C.,  1998, \mn@doi [\araa] {10.1146/annurev.astro.36.1.189},
  \href
  {http://adsabs.harvard.edu/cgi-bin/nph-bib_query?bibcode=1998ARA%26A..36..189K&db_key=AST}
  {36, 189}

\bibitem[\protect\citeauthoryear{{Kennicutt} \& {Evans}}{{Kennicutt} \&
  {Evans}}{2012}]{kennicutt12}
{Kennicutt} R.~C.,  {Evans} N.~J.,  2012, \mn@doi [\araa]
  {10.1146/annurev-astro-081811-125610}, \href
  {http://adsabs.harvard.edu/abs/2012ARA%26A..50..531K} {50, 531}

\bibitem[\protect\citeauthoryear{{Konstantopoulos} et~al.,}{{Konstantopoulos}
  et~al.}{2013}]{konstantopoulos13}
{Konstantopoulos} I.~S.,  et~al., 2013, \mn@doi [\aj]
  {10.1088/0004-6256/145/5/137}, \href
  {http://adsabs.harvard.edu/abs/2013AJ....145..137K} {145, 137}

\bibitem[\protect\citeauthoryear{{Kravtsov} \& {Gnedin}}{{Kravtsov} \&
  {Gnedin}}{2005}]{kravtsov05}
{Kravtsov} A.~V.,  {Gnedin} O.~Y.,  2005, \mn@doi [\apj] {10.1086/428636},
  \href
  {http://adsabs.harvard.edu/cgi-bin/nph-bib_query?bibcode=2005ApJ...623..650K&db_key=AST}
  {623, 650}

\bibitem[\protect\citeauthoryear{{Kruijssen}}{{Kruijssen}}{2012}]{kruijssen12d}
{Kruijssen} J.~M.~D.,  2012, \mn@doi [\mnras]
  {10.1111/j.1365-2966.2012.21923.x}, \href
  {http://adsabs.harvard.edu/abs/2012MNRAS.426.3008K} {426, 3008}

\bibitem[\protect\citeauthoryear{{Kruijssen}}{{Kruijssen}}{2014}]{kruijssen14c}
{Kruijssen} J.~M.~D.,  2014, \mn@doi [Classical and Quantum Gravity]
  {10.1088/0264-9381/31/24/244006}, \href
  {http://adsabs.harvard.edu/abs/2014CQGra..31x4006K} {31, 244006}

\bibitem[\protect\citeauthoryear{{Kruijssen}}{{Kruijssen}}{2015}]{kruijssen15b}
{Kruijssen} J.~M.~D.,  2015, \mn@doi [\mnras] {10.1093/mnras/stv2026}, \href
  {http://adsabs.harvard.edu/abs/2015MNRAS.454.1658K} {454, 1658}

\bibitem[\protect\citeauthoryear{{Kruijssen} \& {Portegies Zwart}}{{Kruijssen}
  \& {Portegies Zwart}}{2009}]{kruijssen09b}
{Kruijssen} J.~M.~D.,  {Portegies Zwart} S.~F.,  2009, \mn@doi [\apjl]
  {10.1088/0004-637X/698/2/L158}, \href
  {http://adsabs.harvard.edu/abs/2009ApJ...698L.158K} {698, L158}

\bibitem[\protect\citeauthoryear{{Kruijssen}, {Longmore}, {Elmegreen},
  {Murray}, {Bally}, {Testi}  \& {Kennicutt}}{{Kruijssen}
  et~al.}{2014}]{kruijssen14b}
{Kruijssen} J.~M.~D.,  {Longmore} S.~N.,  {Elmegreen} B.~G.,  {Murray} N.,
  {Bally} J.,  {Testi} L.,   {Kennicutt} R.~C.,  2014, \mn@doi [\mnras]
  {10.1093/mnras/stu494}, \href
  {http://adsabs.harvard.edu/abs/2014MNRAS.440.3370K} {440, 3370}

\bibitem[\protect\citeauthoryear{{Kruijssen}, {Dale}  \&
  {Longmore}}{{Kruijssen} et~al.}{2015}]{kruijssen15}
{Kruijssen} J.~M.~D.,  {Dale} J.~E.,   {Longmore} S.~N.,  2015, \mn@doi
  [\mnras] {10.1093/mnras/stu2526}, \href
  {http://adsabs.harvard.edu/abs/2015MNRAS.447.1059K} {447, 1059}

\bibitem[\protect\citeauthoryear{{Krumholz} \& {McKee}}{{Krumholz} \&
  {McKee}}{2005}]{krumholz05}
{Krumholz} M.~R.,  {McKee} C.~F.,  2005, \mn@doi [\apj] {10.1086/431734}, \href
  {http://adsabs.harvard.edu/cgi-bin/nph-bib_query?bibcode=2005ApJ...630..250K&db_key=AST}
  {630, 250}

\bibitem[\protect\citeauthoryear{{Krumholz}, {Dekel}  \& {McKee}}{{Krumholz}
  et~al.}{2012}]{krumholz12}
{Krumholz} M.~R.,  {Dekel} A.,   {McKee} C.~F.,  2012, \mn@doi [\apj]
  {10.1088/0004-637X/745/1/69}, \href
  {http://adsabs.harvard.edu/abs/2012ApJ...745...69K} {745, 69}

\bibitem[\protect\citeauthoryear{{Krumholz}, {Kruijssen}  \&
  {Crocker}}{{Krumholz} et~al.}{2017}]{krumholz17}
{Krumholz} M.~R.,  {Kruijssen} J.~M.~D.,   {Crocker} R.~M.,  2017, \mn@doi
  [\mnras] {10.1093/mnras/stw3195}, \href
  {http://adsabs.harvard.edu/abs/2017MNRAS.466.1213K} {466, 1213}

\bibitem[\protect\citeauthoryear{{Lada} \& {Lada}}{{Lada} \&
  {Lada}}{2003}]{lada03}
{Lada} C.~J.,  {Lada} E.~A.,  2003, \mn@doi [\araa]
  {10.1146/annurev.astro.41.011802.094844}, \href
  {http://adsabs.harvard.edu/cgi-bin/nph-bib_query?bibcode=2003ARA%26A..41...57L&db_key=AST}
  {41, 57}

\bibitem[\protect\citeauthoryear{{Larsen}}{{Larsen}}{2009}]{larsen09}
{Larsen} S.~S.,  2009, \mn@doi [\aap] {10.1051/0004-6361:200811212}, \href
  {http://adsabs.harvard.edu/abs/2009A%26A...494..539L} {494, 539}

\bibitem[\protect\citeauthoryear{{Larson}}{{Larson}}{1981}]{larson81}
{Larson} R.~B.,  1981, \mnras, \href
  {http://adsabs.harvard.edu/abs/1981MNRAS.194..809L} {194, 809}

\bibitem[\protect\citeauthoryear{{Leroy}, {Walter}, {Brinks}, {Bigiel}, {de
  Blok}, {Madore}  \& {Thornley}}{{Leroy} et~al.}{2008}]{leroy08}
{Leroy} A.~K.,  {Walter} F.,  {Brinks} E.,  {Bigiel} F.,  {de Blok} W.~J.~G.,
  {Madore} B.,   {Thornley} M.~D.,  2008, \mn@doi [\aj]
  {10.1088/0004-6256/136/6/2782}, \href
  {http://adsabs.harvard.edu/abs/2008AJ....136.2782L} {136, 2782}

\bibitem[\protect\citeauthoryear{{Leroy} et~al.,}{{Leroy}
  et~al.}{2013}]{leroy13}
{Leroy} A.~K.,  et~al., 2013, \mn@doi [\aj] {10.1088/0004-6256/146/2/19}, \href
  {http://adsabs.harvard.edu/abs/2013AJ....146...19L} {146, 19}

\bibitem[\protect\citeauthoryear{{Leroy} et~al.,}{{Leroy}
  et~al.}{2016}]{leroy16}
{Leroy} A.~K.,  et~al., 2016, \mn@doi [\apj] {10.3847/0004-637X/831/1/16},
  \href {http://adsabs.harvard.edu/abs/2016ApJ...831...16L} {831, 16}

\bibitem[\protect\citeauthoryear{{Lewis} et~al.,}{{Lewis}
  et~al.}{2015}]{lewis15}
{Lewis} A.~R.,  et~al., 2015, \mn@doi [\apj] {10.1088/0004-637X/805/2/183},
  \href {http://adsabs.harvard.edu/abs/2015ApJ...805..183L} {805, 183}

\bibitem[\protect\citeauthoryear{{Longmore} et~al.,}{{Longmore}
  et~al.}{2012}]{longmore12}
{Longmore} S.~N.,  et~al., 2012, \mn@doi [\apj] {10.1088/0004-637X/746/2/117},
  \href {http://adsabs.harvard.edu/abs/2012ApJ...746..117L} {746, 117}

\bibitem[\protect\citeauthoryear{{Longmore} et~al.,}{{Longmore}
  et~al.}{2013}]{longmore13}
{Longmore} S.~N.,  et~al., 2013, \mn@doi [\mnras] {10.1093/mnras/sts376}, \href
  {http://adsabs.harvard.edu/abs/2013MNRAS.429..987L} {429, 987}

\bibitem[\protect\citeauthoryear{{Longmore} et~al.,}{{Longmore}
  et~al.}{2014}]{longmore14}
{Longmore} S.~N.,  et~al., 2014, \mn@doi [Protostars and Planets VI]
  {10.2458/azu_uapress_9780816531240-ch013}, \href
  {http://adsabs.harvard.edu/abs/2014prpl.conf..291L} {pp 291--314}

\bibitem[\protect\citeauthoryear{{Mancini} et~al.,}{{Mancini}
  et~al.}{2011}]{mancini11}
{Mancini} C.,  et~al., 2011, \mn@doi [\apj] {10.1088/0004-637X/743/1/86}, \href
  {http://adsabs.harvard.edu/abs/2011ApJ...743...86M} {743, 86}

\bibitem[\protect\citeauthoryear{{Milam}, {Savage}, {Brewster}, {Ziurys}  \&
  {Wyckoff}}{{Milam} et~al.}{2005}]{milam05}
{Milam} S.~N.,  {Savage} C.,  {Brewster} M.~A.,  {Ziurys} L.~M.,   {Wyckoff}
  S.,  2005, \mn@doi [\apj] {10.1086/497123}, \href
  {http://adsabs.harvard.edu/abs/2005ApJ...634.1126M} {634, 1126}

\bibitem[\protect\citeauthoryear{{Miville-Desch{\^e}nes}, {Murray}  \&
  {Lee}}{{Miville-Desch{\^e}nes} et~al.}{2017}]{mivilledeschenes17}
{Miville-Desch{\^e}nes} M.-A.,  {Murray} N.,   {Lee} E.~J.,  2017, \mn@doi
  [\apj] {10.3847/1538-4357/834/1/57}, \href
  {http://adsabs.harvard.edu/abs/2017ApJ...834...57M} {834, 57}

\bibitem[\protect\citeauthoryear{{Moster}, {Naab}  \& {White}}{{Moster}
  et~al.}{2013}]{moster13}
{Moster} B.~P.,  {Naab} T.,   {White} S.~D.~M.,  2013, \mn@doi [\mnras]
  {10.1093/mnras/sts261}, \href
  {http://adsabs.harvard.edu/abs/2013MNRAS.428.3121M} {428, 3121}

\bibitem[\protect\citeauthoryear{{Murray}}{{Murray}}{2011}]{murray11}
{Murray} N.,  2011, \mn@doi [\apj] {10.1088/0004-637X/729/2/133}, \href
  {http://adsabs.harvard.edu/abs/2011ApJ...729..133M} {729, 133}

\bibitem[\protect\citeauthoryear{{Oklop{\v c}i{\'c}}, {Hopkins}, {Feldmann},
  {Kere{\v s}}, {Faucher-Gigu{\`e}re}  \& {Murray}}{{Oklop{\v c}i{\'c}}
  et~al.}{2017}]{oklopcic17}
{Oklop{\v c}i{\'c}} A.,  {Hopkins} P.~F.,  {Feldmann} R.,  {Kere{\v s}} D.,
  {Faucher-Gigu{\`e}re} C.-A.,   {Murray} N.,  2017, \mn@doi [\mnras]
  {10.1093/mnras/stw2754}, \href
  {http://adsabs.harvard.edu/abs/2017MNRAS.465..952O} {465, 952}

\bibitem[\protect\citeauthoryear{{Oort} \& {Spitzer}}{{Oort} \&
  {Spitzer}}{1955}]{oort55}
{Oort} J.~H.,  {Spitzer} Jr. L.,  1955, \mn@doi [\apj] {10.1086/145958}, \href
  {http://adsabs.harvard.edu/abs/1955ApJ...121....6O} {121, 6}

\bibitem[\protect\citeauthoryear{{Portegies Zwart}, {McMillan}  \&
  {Gieles}}{{Portegies Zwart} et~al.}{2010}]{portegieszwart10}
{Portegies Zwart} S.~F.,  {McMillan} S.~L.~W.,   {Gieles} M.,  2010, \mn@doi
  [\araa] {10.1146/annurev-astro-081309-130834}, \href
  {http://adsabs.harvard.edu/abs/2010ARA%26A..48..431P} {48, 431}

\bibitem[\protect\citeauthoryear{{Prieto} \& {Gnedin}}{{Prieto} \&
  {Gnedin}}{2008}]{prieto08}
{Prieto} J.~L.,  {Gnedin} O.~Y.,  2008, \mn@doi [\apj] {10.1086/591777}, \href
  {http://adsabs.harvard.edu/abs/2008ApJ...689..919P} {689, 919}

\bibitem[\protect\citeauthoryear{{Rieder}, {Ishiyama}, {Langelaan}, {Makino},
  {McMillan}  \& {Portegies Zwart}}{{Rieder} et~al.}{2013}]{rieder13}
{Rieder} S.,  {Ishiyama} T.,  {Langelaan} P.,  {Makino} J.,  {McMillan}
  S.~L.~W.,   {Portegies Zwart} S.,  2013, \mn@doi [\mnras]
  {10.1093/mnras/stt1848}, \href
  {http://adsabs.harvard.edu/abs/2013MNRAS.436.3695R} {436, 3695}

\bibitem[\protect\citeauthoryear{{Schechter}}{{Schechter}}{1976}]{schechter76}
{Schechter} P.,  1976, \apj, \href
  {http://adsabs.harvard.edu/cgi-bin/nph-bib_query?bibcode=1976ApJ...203..297S&db_key=AST}
  {203, 297}

\bibitem[\protect\citeauthoryear{{Schruba} et~al.,}{{Schruba}
  et~al.}{2011}]{schruba11}
{Schruba} A.,  et~al., 2011, \mn@doi [\aj] {10.1088/0004-6256/142/2/37}, \href
  {http://cdsads.u-strasbg.fr/abs/2011AJ....142...37S} {142, 37}

\bibitem[\protect\citeauthoryear{{Schruba} et~al.,}{{Schruba}
  et~al.}{2017}]{schruba17}
{Schruba} A.,  et~al., 2017, in prep.

\bibitem[\protect\citeauthoryear{{Shapiro}, {Genzel}  \& {F{\"o}rster
  Schreiber}}{{Shapiro} et~al.}{2010}]{shapiro10}
{Shapiro} K.~L.,  {Genzel} R.,   {F{\"o}rster Schreiber} N.~M.,  2010, \mn@doi
  [\mnras] {10.1111/j.1745-3933.2010.00810.x}, \href
  {http://adsabs.harvard.edu/abs/2010MNRAS.403L..36S} {403, L36}

\bibitem[\protect\citeauthoryear{{Solomon}, {Rivolo}, {Barrett}  \&
  {Yahil}}{{Solomon} et~al.}{1987}]{solomon87}
{Solomon} P.~M.,  {Rivolo} A.~R.,  {Barrett} J.,   {Yahil} A.,  1987, \mn@doi
  [\apj] {10.1086/165493}, \href
  {http://adsabs.harvard.edu/abs/1987ApJ...319..730S} {319, 730}

\bibitem[\protect\citeauthoryear{{Soto} et~al.,}{{Soto} et~al.}{2017}]{soto17}
{Soto} E.,  et~al., 2017, \apj~in~press, arXiv:1702.03038, \href
  {http://adsabs.harvard.edu/abs/2017arXiv170203038S} {}

\bibitem[\protect\citeauthoryear{{Tacconi} et~al.,}{{Tacconi}
  et~al.}{2010}]{tacconi10}
{Tacconi} L.~J.,  et~al., 2010, \mn@doi [\nat] {10.1038/nature08773}, \href
  {http://adsabs.harvard.edu/abs/2010Natur.463..781T} {463, 781}

\bibitem[\protect\citeauthoryear{{Tacconi} et~al.,}{{Tacconi}
  et~al.}{2013}]{tacconi13}
{Tacconi} L.~J.,  et~al., 2013, \mn@doi [\apj] {10.1088/0004-637X/768/1/74},
  \href {http://adsabs.harvard.edu/abs/2013ApJ...768...74T} {768, 74}

\bibitem[\protect\citeauthoryear{{Tamburello}, {Mayer}, {Shen}  \&
  {Wadsley}}{{Tamburello} et~al.}{2015}]{tamburello15}
{Tamburello} V.,  {Mayer} L.,  {Shen} S.,   {Wadsley} J.,  2015, \mn@doi
  [\mnras] {10.1093/mnras/stv1695}, \href
  {http://adsabs.harvard.edu/abs/2015MNRAS.453.2490T} {453, 2490}

\bibitem[\protect\citeauthoryear{{Toomre}}{{Toomre}}{1964}]{toomre64}
{Toomre} A.,  1964, \mn@doi [\apj] {10.1086/147861}, \href
  {http://adsabs.harvard.edu/abs/1964ApJ...139.1217T} {139, 1217}

\bibitem[\protect\citeauthoryear{{Vansevi{\v c}ius}, {Kodaira}, {Narbutis},
  {Stonkut{\.e}}, {Brid{\v z}ius}, {Deveikis}  \& {Semionov}}{{Vansevi{\v
  c}ius} et~al.}{2009}]{vansevicius09}
{Vansevi{\v c}ius} V.,  {Kodaira} K.,  {Narbutis} D.,  {Stonkut{\.e}} R.,
  {Brid{\v z}ius} A.,  {Deveikis} V.,   {Semionov} D.,  2009, \mn@doi [\apj]
  {10.1088/0004-637X/703/2/1872}, \href
  {http://adsabs.harvard.edu/abs/2009ApJ...703.1872V} {703, 1872}

\bibitem[\protect\citeauthoryear{{Walker}, {Longmore}, {Bastian}, {Kruijssen},
  {Rathborne}, {Jackson}, {Foster}  \& {Contreras}}{{Walker}
  et~al.}{2015}]{walker15}
{Walker} D.~L.,  {Longmore} S.~N.,  {Bastian} N.,  {Kruijssen} J.~M.~D.,
  {Rathborne} J.~M.,  {Jackson} J.~M.,  {Foster} J.~B.,   {Contreras} Y.,
  2015, \mn@doi [\mnras] {10.1093/mnras/stv300}, \href
  {http://adsabs.harvard.edu/abs/2015MNRAS.449..715W} {449, 715}

\bibitem[\protect\citeauthoryear{{Walter}, {Brinks}, {de Blok}, {Bigiel},
  {Kennicutt}, {Thornley}  \& {Leroy}}{{Walter} et~al.}{2008}]{walter08}
{Walter} F.,  {Brinks} E.,  {de Blok} W.~J.~G.,  {Bigiel} F.,  {Kennicutt}
  R.~C.~J.,  {Thornley} M.~D.,   {Leroy} A.,  2008, \mn@doi [\aj]
  {10.1088/0004-6256/136/6/2563}, \href
  {http://adsabs.harvard.edu/abs/2008AJ....136.2563W} {136, 2563}

\makeatother
\end{thebibliography}

\bsp

\label{lastpage}

\end{document}